\documentclass[preprint,12pt]{elsarticle}

\providecommand{\Description}[1]{}

\usepackage{amssymb}
\usepackage{amsmath}
\usepackage{tabularx}
\usepackage{graphicx}
\usepackage{cancel}
\usepackage{subcaption}
\usepackage{array}
\usepackage{booktabs}
\usepackage[ruled,vlined]{algorithm2e}
\usepackage{float}

\begin{document}

\begin{frontmatter}

\title{SGEMM-cube: Precision-Recovery FP32 GEMM Approximation on Ascend NPUs with FP16 Matrix Engines}

\author[inst1]{Weicheng Xue\fnref{equal}}
\author[inst1]{Baisong Xu\fnref{equal}}
\author[inst1]{Kai Yang\fnref{equal}}
\author[inst1]{Yongxiang Liu}
\author[inst1]{Dengdeng Fan}
\author[inst1]{Pengxiang Xu\corref{cor1}}
\author[inst1]{Yonghong Tian\corref{cor1}}

\affiliation[inst1]{organization={Pengcheng Laboratory},
addressline={6001 Shahe West Road, Shibilong},
city={Shenzhen},
postcode={518055},
state={Guangdong},
country={China}}

\fntext[equal]{These authors contributed equally to this work.}
\cortext[cor1]{Corresponding authors: Pengxiang Xu (xupx@pcl.ac.cn), Yonghong Tian (tianyh@pcl.ac.cn).}

\begin{abstract}
Modern AI accelerators provide high-throughput low-precision matrix engines, but often lack efficient support for FP32 GEMM. This paper presents SGEMM-cube, an FP32-accuracy GEMM approximation for Ascend NPUs built on FP16 Cube units. Following the fixed-length two-word splitting of Ootomo and Yokota, each FP32 operand is represented by an FP16 high component and a scaled FP16 residual. The product is reconstructed from three FP16 GEMMs while omitting the residual-residual term; the method therefore targets FP32-level accuracy rather than bit-exact emulation. Under an RN FP32-accumulation model, we provide a componentwise error
analysis showing that the omitted term is no larger than the rounding error of a short FP32 inner product and that, for practically relevant inner-product lengths, the overall error is dominated by ordinary FP32 accumulation. We further analyze residual underflow and scaling under round-to-nearest conversion, compare two accumulation orders, and adapt L1-aware blocking and double buffering to Ascend's software-managed memory hierarchy. On Ascend 910A, SGEMM-cube is substantially more accurate than native FP16 GEMM, is comparable to the tested OpenBLAS FP32 SGEMM baseline for the evaluated input distributions and exponent range, and reaches 65.3\,TFLOP/s, or 77\% of the three-GEMM FP32-equivalent peak.
\end{abstract}


\begin{highlights}
\item Present an Ascend NPU implementation of the Ootomo-style two-word FP32-accuracy GEMM approximation using FP16 Cube units.
\item Under an RN FP32-accumulation model, derive a componentwise
error bound for the three-term reconstruction that quantifies the
omitted low-low product.
\item Analyze residual scaling under round-to-nearest FP32-to-FP16 conversion and clarify the resulting accuracy and range limitations.
\item Compare elementwise and termwise accumulation strategies and show their impact on numerical stability.
\item Adapt standard GEMM blocking and double-buffering techniques to Ascend's software-managed memory hierarchy.
\item Provide accuracy and performance evaluation on Ascend 910A, achieving up to 65.3 TFLOP/s, or 77\% of the three-GEMM FP32-equivalent peak.
\end{highlights}


\end{frontmatter}

\section{Introduction}

The evolution of specialized accelerators is reshaping the landscape of high-performance computing (HPC). Driven by the rapid progress in deep learning, modern computer systems increasingly rely on heterogeneous architectures that integrate domain-specific hardware. Processors such as NVIDIA GPUs with Tensor Cores, AMD GPUs with Matrix Cores~\cite{ambati2025amd}, Intel's Gaudi accelerators~\cite{kaplan2024intel}, and emerging Ascend Neural Processing Units (NPUs)~\cite{liao2021ascend} deliver significant speedups for matrix-dominated workloads, offering orders of magnitude higher throughput than traditional CPUs.

These accelerators achieve high performance by concentrating their compute resources on low-precision formats such as half-precision floating point (FP16), bfloat16 (BF16), and INT8. Though effective for many deep-learning workloads, low-precision formats often fail to meet the accuracy and dynamic range requirements of scientific computing and of numerically sensitive machine-learning kernels~\cite{yu2021any,zhuang2020training, vansteenkiste2014design, muller2019spirit, wozniak2016gimmik, cawkwell2012computing}, which typically require FP32 or higher precision. This discrepancy creates a major challenge: state-of-the-art accelerators are tuned for low-precision formats that cannot satisfy the precision needs of many scientific workloads.

To bridge this gap, this work investigates how to efficiently harness the compute density of low-precision hardware for workloads that require FP32-level accuracy. Specifically, we target scenarios where native FP32 matrix multiplication (GEMM) is unsupported or inefficient, such as on Ascend 910A NPUs that expose high-performance FP16 matrix engines but lack equivalent FP32 compute units. We present SGEMM-cube, an Ascend-specific implementation and analysis of an Ootomo-style two-word FP32-accuracy GEMM approximation that uses FP16 Cube units for the throughput-dominant matrix multiplications and FP32 arithmetic for reconstruction and accumulation.

Some terminology needs to be fixed before going further, since two distinctions are routinely blurred in this area. Emulating a high-precision operation with low-precision arithmetic is not the same as running a mixed-precision algorithm. Mixed-precision iterative refinement~\cite{haidar2018harnessing, higham2022mixed} executes different operations of an algorithm in different formats and recovers accuracy through the structure of the algorithm; emulation reproduces one high-precision operation, here a single GEMM, using only low-precision units. Only the latter problem is treated here. The second distinction is between the Ozaki scheme~\cite{ozaki2012error} and the fixed-length multiword splitting of Ootomo and Yokota~\cite{ootomo2022recovering}. These are frequently referred to interchangeably, but they differ in an essential way, and it is the latter that SGEMM-cube implements. Sec.~\ref{sec:ozaki_vs_ootomo} sets out the difference.

Existing implementations are also not directly transferable. Solutions developed for NVIDIA GPUs and their Tensor Cores assume a relatively mature ecosystem: hardware-managed caches, a well-understood shared-memory hierarchy, and the CUDA programming model~\cite{ootomo2022recovering, ootomo2024dgemm}. The Ascend NPU exposes an L1/L0A/L0B/L0C/UB hierarchy that is managed entirely in software. Every transfer is issued explicitly, and no hardware cache masks a poor blocking choice. Its machine balance is high as well: 256\,TFLOP/s of FP16 Cube throughput against 1.2\,TB/s of HBM bandwidth gives $\approx\!213$ FLOP/Byte, against $\approx\!153$ FLOP/Byte for an NVIDIA A100 SXM (312\,TFLOP/s FP16 tensor, 2.04\,TB/s), so a GEMM must sustain roughly $1.4\times$ more reuse to stay compute-bound~\cite{gholami2024ai}. Schedules tuned for a GPU memory model do not port efficiently, which leaves an open question about how well precision-recovery techniques carry across accelerator families. We approach it by tuning the numerical formulation and the blocking and pipelining schedule together for a bandwidth-constrained, software-managed NPU.

At its core, SGEMM-cube combines a numerical approximation technique with a performance-optimized implementation tailored to the memory hierarchy of a representative AI accelerator. The contributions of this work are as follows:

\begin{itemize}
  \item \textbf{Accuracy and range envelope:}
    The two-word FP32-to-FP16 split is implemented on Ascend NPUs and characterized as an FP32-\emph{accuracy approximation}, not a bit-exact FP32 emulation. The 22-bit relative-error bound applies when the FP16 high component remains normal and the scaled residual neither underflows nor overflows; under these conditions, the representation has an effective unit roundoff of $2^{-22}$, compared with $2^{-24}$ for FP32.

  \item \textbf{Componentwise error bound for the three-term reconstruction:}
  Under an RN FP32-accumulation model, Sec.~\ref{sec:err_bound} derives a conservative entrywise error bound for the termwise reconstruction that separates representation, truncation, and FP32 accumulation errors. The exact low-low coefficient has the same leading-order scale as a length-four FP32 accumulation bound and is smaller than the standard FP32 accumulation bound for $k\geq 5$. Retaining the higher-order terms, the complete bound is within a factor of two of the classical FP32 GEMM bound for $k\geq 15$.

  \item \textbf{RN-based scaling and range analysis:}
  Residual underflow and overflow are analyzed under round-to-nearest conversion, where earlier work assumed round-toward-zero. The analysis gives explicit lower and upper bounds on the scaling exponent. We use $2^{12}$ for the evaluated low-magnitude exponent window, while a fixed scale that covers the full normal-FP16 high-component range must use $2^{11}$ or an explicit overflow guard.

  \item \textbf{Accumulation order:}
  Elementwise and termwise reconstruction are compared. Termwise accumulation aggregates the correction terms before combining them with the high-order product, and is measurably more stable in the low-exponent regime.

  \item \textbf{Implementation and measurement:}
  Blocking and double buffering, both standard techniques, are adapted to the software-managed memory hierarchy of Ascend. We measure what each of them contributes to throughput, and compare the result with FP16 HGEMM and FP32 baselines.
\end{itemize}

Much of the above builds directly on existing results, so we should say at the outset what belongs to whom. The splitting formula, the omission of the low-low term, and the sign flip induced by round-to-nearest conversion are due to Ootomo and Yokota~\cite{ootomo2022recovering}. The general error analysis of multiword products is due to Fasi et al.~\cite{fasi2023multiword}, and its narrow-range counterpart to Mary and Mikaitis~\cite{mary2025narrow}. The blocking and double-buffering of Sec.~\ref{system_optimizations} follow Goto and van de Geijn~\cite{goto2008anatomy}. Ours is the specialization to the FP16/FP32 pair on an FP16-only, software-managed matrix engine, the explicit constants and scaling rules that come out of it, and the measurements on Ascend hardware.

\section{Background and Related Work}

The increasing divergence between the numerical precision demands of scientific computing and the capabilities of modern AI-oriented accelerators (seen in Table~\ref{chip_flops}) has spurred significant research into mixed-precision algorithms~\cite{micikevicius2017mixed, rakka2022mixed, haidar2018harnessing, higham2022mixed}. While many scientific applications still demand FP32 or FP64 precision for numerical stability, contemporary hardware increasingly favors lower-precision formats such as FP16 or even FP8 to maximize throughput and energy efficiency. For example, the FP8 Tensor Core throughput of the NVIDIA H100 is roughly 60$\times$ higher than its FP64 throughput, and the Ascend 910A NPU~\cite{liao2021ascend} delivers $\sim$ 256 TF/s in FP16 but lacks efficient support for FP32 GEMM. These trends motivate a practical question: \textit{how much FP32 accuracy can be recovered from low-precision matrix engines, and under what range and implementation constraints?}\label{sec:background_question} We answer the accuracy half of this question with the rounding-error bound of Sec.~\ref{sec:err_bound} and the range half with the underflow analysis of Sec.~\ref{sec:accuracy_analysis}.

\begin{table}[!t]
\centering
\caption{Peak throughput of representative AI accelerators in TFLOP/s}
\footnotesize
\begin{tabular}{lccc}
\toprule
Chip Model & FP16 & FP32 & FP64 \\\hline
Nvidia H100 SXM & 989 & 67 & 34 \\
Nvidia A100 SXM & 312 & 19.5 & 9.7 \\
AMD MI300X~\cite{ambati2025amd} & 1307 & 163 & 81 \\
Intel Gaudi 3~\cite{kaplan2024intel} & 459 & 229 & - \\
Huawei Ascend 910A~\cite{liao2021ascend} & 256 & - & - \\
Cambricon MLU370-X8 & 96 & 24 & - \\
Muxi Xiyun C500 & 280 & 36 & - \\
Shenwei SW26010-Pro~\cite{10363670} & 55.3 & 14.0 & 14.0 \\
Moore Threads MTT S4000 & 100 & 25 & - \\
\bottomrule
\end{tabular}
\label{chip_flops}
\end{table}

\subsection{Mixed-precision algorithms versus emulation}
\label{sec:mp_vs_emulation}

Two research directions are often grouped together under the heading of low-precision arithmetic, although they solve different problems. A mixed-precision algorithm executes different operations of a single algorithm in different formats. Mixed-precision iterative refinement, for instance, computes an LU factorization in low precision and refines the solution using residuals evaluated in high precision~\cite{haidar2018harnessing, higham2022mixed, micikevicius2017mixed, rakka2022mixed}. Accuracy in the final output is recovered by the structure of the algorithm, not by any individual kernel. Emulation, or precision recovery, instead approximates one high-precision operation using low-precision units. SGEMM-cube belongs to this second category, but only within the input-range and scaling constraints stated in Sec.~\ref{method}; it is not a general drop-in replacement for IEEE-754 FP32 GEMM.

\subsection{Error-free (Ozaki) splitting versus fixed-length multiword (Ootomo) splitting}
\label{sec:ozaki_vs_ootomo}

Emulation itself splits into two families, and the present work belongs to the second of them.

The Ozaki scheme~\cite{ozaki2012error, mukunoki2020dgemm, ootomo2024dgemm, uchino2025ozaki} is an error-free transformation. Each operand is split into a data-dependent number of slices, with exponent ranges chosen so that every pairwise slice product and its accumulation are exact in the working precision. The full product is recovered as a sum of exactly computed low-precision GEMMs. Increasing the number of splits improves accuracy until only the final summation limits it, and in the limit the exact product can be correctly rounded. Cost is the price of this generality: it grows with the dynamic range of the inputs and with the requested accuracy, and the number of GEMMs is not known in advance.

The Ootomo-style scheme~\cite{markidis2018nvidia, feng2021egemm, ootomo2022recovering, ma2022efficiently} uses a fixed-length multiword representation instead. Each operand is written as a fixed number of low-precision words, two in our case, and the product is approximated by keeping only the leading cross terms of the expansion. The discarded terms and the truncation of the representation both contribute a nonzero, though bounded, error, so this is an approximation and not an error-free transformation. In exchange, the cost is fixed and known: three FP16 GEMMs for a two-word split with the low-low term dropped. That predictability is what makes the scheme attractive on a throughput-oriented matrix engine.

We avoid the label ``Ozaki scheme'' for SGEMM-cube throughout, and quantify the resulting approximation error in Sec.~\ref{sec:err_bound}.

\subsection{FP32-from-FP16 precision recovery on matrix engines}

Markidis et al.~\cite{markidis2018nvidia} introduced residual-refinement schemes for Tensor Core GEMM: a one-sided variant using two GEMMs and a two-sided variant using four GEMMs. These variants reduce the error caused by converting FP32 inputs to FP16 at additional computation and storage cost. Ootomo and Yokota~\cite{ootomo2022recovering} later identified
internal Tensor Core rounding and correction-term underflow as major remaining obstacles to matching FP32 SIMT-core accuracy.

To reduce this error, Feng et al.~\cite{feng2021egemm} proposed Egemm-TC, a refinement that adopted a more careful decomposition and a rounding strategy closer to round-to-nearest (RN). Their implementation included warp-level cache reuse and register pipeline optimizations, yielding a 3.13$\times$ speedup over cuBLAS FP32 GEMM on NVIDIA Turing GPUs. However, the decomposition still failed to fully account for the implicit leading bit in FP32 representation, limiting the worst-case precision to approximately 21 bits.

Ootomo and Yokota~\cite{ootomo2022recovering} later identified another key source of error: the use of Tensor Cores for accumulation. Tensor Core units perform multiplications in FP16 but may accumulate results using internal FP32 registers with RZ rounding~\cite{fasi2021numerical}. This caused a systematic bias that could not be corrected by earlier schemes. They proposed performing the residual accumulation outside the Tensor Core, using standard FP32 units with unbiased RN rounding. By scaling intermediate terms to prevent underflow and carefully structuring the summation, their implementation achieved accuracy matching FP32 SIMT-core GEMM
in their experiments, reaching 51\,TFLOP/s for a limited exponent range with FP16 Tensor Cores and 33\,TFLOP/s over the full FP32
exponent range with TF32 Tensor Cores. To support matrices spanning the full FP32 exponent range, they later integrated TensorFloat-32 (TF32) into their framework, attaining $\sim$33 TFLOPS with negligible ($\sim$ 1-bit) precision loss.

Ma et al.~\cite{ma2022efficiently} presented APE, a framework that
provides multiple low-precision strategies for FP32 emulation, including FP16-, BF16-, and TF32-based variants, together with error analysis and automatic strategy selection under range constraints. APE was implemented on both NVIDIA Tensor Core GPUs and Huawei Ascend processors. Its design space is broader than the fixed formulation studied here: SGEMM-cube focuses on a two-word FP16, three-GEMM reconstruction on Ascend 910A, whereas APE also includes other representation formats and operation counts; for example, its FP32-B configuration uses six BF16 matrix
multiplications.

In parallel to these FP32-from-FP16 approximation schemes, Li et al.~\cite{li2021unleashing} proposed QuanTensor, a general low-precision tensor computation framework for GPUs. QuanTensor decomposes high-precision GEMMs into multiple low-precision GEMMs combined with residual correction to recover accuracy. By flexibly choosing the low-precision type and number of passes, it enables a tunable trade-off between throughput and precision. On NVIDIA GPUs, the INT8 variant achieves the highest throughput, while FP16/BF16 variants offer better numerical stability with moderate performance gains. However, its multi-pass design requires repeated quantization and dequantization on the GPU platform, which can incur non-trivial overhead when many passes are used or when high accuracy is required.

Lin et al.~\cite{lin2024mixpert} proposed MixPert, a compiler-integrated framework that approximates FP32 GEMM on GPU Integer Tensor Cores (ITCs) by decomposing each operand into four INT8 segments. While prior work relies on FP16 or TF32, MixPert exploits the $8\times\sim60\times$ throughput of ITCs to deliver $1.72\times$ speed-up over cuBLAS FP32 with lower error than native FP32. A lightweight auto-tuner selects one of four accuracy levels under a user-supplied error bound, enabling deployment on both high-end and commodity GPUs without manual tuning.

\subsection{General error analyses of multiword and narrow-range products}
\label{sec:general_analysis}

The schemes above were originally analyzed one at a time. A general theory now exists that subsumes the FP16/FP32 instance studied here. Fasi et al.~\cite{fasi2023multiword} represent a matrix as the unevaluated sum of two or more lower-precision matrices and, on top of a general block-FMA framework, develop an error analysis of multiword matrix multiplication covering the whole family to which the two-word FP16 split belongs. Their experiments on V100 and A100 GPUs also report something directly relevant here: double-fp16 satisfies the same worst-case bound as fp32 yet is measurably less accurate in practice, which they attribute to the rounding mode used inside the tensor cores. Mary and Mikaitis~\cite{mary2025narrow} take up the complementary difficulty, namely that low-precision formats have a narrow exponent range, so that accuracy is limited by underflow and overflow and not by rounding alone. They show that with a suitable scaling the underflow errors stay below the rounding errors. Blanchard et al.~\cite{blanchard2020block} and Fasi et al.~\cite{fasi2021numerical} characterize the block-FMA and internal-rounding behavior of the hardware units themselves, and Higham and Mary~\cite{higham2022mixed} survey the area.

Sec.~\ref{sec:accuracy_analysis} does not generalize any of this. The bound in Sec.~\ref{sec:err_bound} is the two-word FP16/FP32 specialization of~\cite{fasi2023multiword} with the constants written out, and the scaling rules in Sec.~\ref{scaling} instantiate the principle of~\cite{mary2025narrow} for FP16 under round-to-nearest conversion, where several earlier implementation papers assumed round-toward-zero. The reason for writing them out is practical. Whether the scheme is usable for a given workload depends on concrete quantities: the two-word representation retains 22 significand bits, the residual is amplified by $2^{12}$ only within a stated exponent window, the exact low-low coefficient is dominated by the standard FP32 accumulation bound for $k\geq 5$, and the retained complete bound is within a factor of two of the classical FP32 bound for $k\geq 15$. These quantities turn out to predict the behavior we later measure.

\subsection{Positioning of this work}

Prior work is not confined to CUDA: APE~\cite{ma2022efficiently} was also implemented on Huawei Ascend processors. SGEMM-cube therefore does not claim the first precision-recovery GEMM implementation on Ascend. Its scope is formulation-specific: we specialize a fixed two-word FP16, three-GEMM reconstruction to Ascend 910A, derive explicit RN-based range and componentwise error bounds for this formulation, compare termwise and elementwise reconstruction, and evaluate an L1-aware implementation
on the 910A software-managed memory hierarchy.

SGEMM-cube is best understood as an Ascend-specific realization of a known strategy, not as a new splitting algorithm. From Ootomo and Yokota~\cite{ootomo2022recovering} we take the two-word high/residual FP16 representation, the amplification of the residual by a power of two before conversion, the fact that round-to-nearest conversion of the high word can flip the sign of the residual, and the decision to omit the residual-residual product. The error-analysis framework comes from~\cite{fasi2023multiword, mary2025narrow} and the blocking and double-buffering methodology from~\cite{goto2008anatomy}. What we add is a componentwise bound for this particular three-term reconstruction, showing that the exact low-low coefficient is below the standard FP32 accumulation bound for inner products of length five or more and that the retained full error bound stays within a factor of two of the classical FP32 GEMM bound once the inner-product length reaches 15; an underflow and scaling analysis carried out under RN conversion instead of RZ, which uses a residual amplification factor of $2^{12}$ within the evaluated exponent window; a comparison of termwise against elementwise accumulation; and an implementation and measurement campaign on an FP16-only, software-managed matrix engine, with measured blocking and pipeline ablations explaining the achieved 77\% of the three-GEMM FP32-equivalent peak. An earlier Ascend study targeting low-bit HPL-MxP~\cite{xue2024unlocking} indicates that kernels of this kind matter in practice on the same platform.

\begin{table}[ht]
\caption{Comparison of FP32 approximation methods using low-precision matrix engines. \textbf{Note:} For SGEMM-cube, the FP32-equivalent peak is defined as one third of the native FP16 peak, because the method computes one approximate FP32 GEMM using three dominant FP16 GEMM terms.}
\centering
\renewcommand{\arraystretch}{1.15}
\setlength{\tabcolsep}{3pt}
\begin{tabular}{
    >{\tiny\raggedright\arraybackslash}p{1.5cm} 
    >{\tiny\raggedright\arraybackslash}p{1.7cm} 
    >{\tiny\raggedright\arraybackslash}p{2.3cm} 
    >{\tiny\raggedright\arraybackslash}p{1.2cm} 
    >{\tiny\raggedright\arraybackslash}p{2.9cm} 
    >{\tiny\raggedright\arraybackslash}p{2.9cm}
}
\hline
\textbf{Work} & 
\textbf{Hardware Target} & 
\textbf{Decomposition / Rounding Method} & 
\textbf{Precision Loss} & 
\textbf{Error Control Mechanisms} & 
\textbf{Performance Claim} \\
\hline
Markidis et al.~\cite{markidis2018nvidia} &
NVIDIA V100 &
One-sided residual refinement (2 GEMMs) or two-sided refinement (4 GEMMs) &
Input-dependent &
Residual correction on one or both operands &
Accuracy--cost trade-off \\
Feng et al.~\cite{feng2021egemm} & NVIDIA T4 \& RTX6000 & Decomposition without hidden bit, RZ & 2 bits & Warp-level cache/register pipeline optimization & 3.13$\times$ speedup over cuBLAS FP32 \\
Ootomo et al.~\cite{ootomo2022recovering} & NVIDIA A100 & Decomposition with amplification, RN & 1 bit & Accumulation outside Tensor Core to bypass RZ rounding & 51 TFLOPS (surpassing theoretical FP32 peak) \\
Ma et al.~\cite{ma2022efficiently} &
NVIDIA V100, T4, A100; Huawei Ascend &
APE: FP16/BF16/TF32 emulation with automatic strategy selection &
Strategy-dependent &
Error analysis; overflow/underflow-aware strategy selection &
64.15 TFLOPS on A100 \\
Li et al.~\cite{li2021unleashing} & NVIDIA T4, RTX2080Ti & Multi-pass low-precision GEMM (FP16/INT8/INT4) with residual correction & \text{N/A} & Configurable \& Flexible decomposition & Tunable precision–performance trade-off \\
Lin et al.~\cite{lin2024mixpert} & NVIDIA A100 & INT8 fixed-point decomposition, RN & 3 bits & Compiler auto-tuning, exponent alignment & 1.72$\times$ over cuBLAS FP32 \\
\textbf{SGEMM-cube} &
\textbf{Ascend 910A} &
\textbf{Ootomo-style two-word FP16 split with RN-based residual scaling} &
\textbf{2 significand bits in the worst case, range-dependent} &
\textbf{Componentwise error bound; RN scaling rules; termwise accumulation} &
\textbf{65.3 TFLOPS, 77\% of three-GEMM FP32-equivalent peak} \\
\hline
\end{tabular}
\label{tab:fp32_emulation}
\end{table}

\section{Problem Scope and Precision-Recovery Formulation}

\label{method}

\subsection{Scope and limitations}
SGEMM-cube does not provide bit-exact FP32 emulation over the full IEEE-754 FP32 dynamic range. Instead, it provides an FP32-accuracy approximation for matrix inputs whose magnitudes can be represented through FP16 high components and scaled FP16 residual components. In particular, the high component of each operand is stored in FP16 and is therefore limited by the FP16 normal/subnormal range. Inputs larger than the FP16 maximum value may overflow during the high-component conversion unless additional exponent management is introduced. Similarly, inputs far below the FP16 subnormal range cannot be accurately represented unless both the high and low components are scaled.

Therefore, SGEMM-cube targets moderate-range FP32 inputs, which commonly occur in many machine-learning and scientific workloads after normalization or problem-specific scaling. The method should not be interpreted as a full replacement for IEEE-754 FP32 arithmetic. It uses FP16 Cube units for the throughput-dominant GEMM operations, but still requires FP32 accumulation, FP32 storage, and FP32 scalar/vector arithmetic for reconstruction.

\subsection{Notation}
\label{sec:notation}

The argument below moves between bit-level, matrix-level and blocking-level descriptions. Table~\ref{tab:notation} collects the symbols used at all three levels; every symbol appearing later in the paper is listed there.

\begin{table}[!p]
\caption{Notation used throughout the paper.}
\label{tab:notation}
\centering
\footnotesize
\begin{tabular}{@{}ll@{}}
\toprule
Symbol & Meaning \\
\midrule
\multicolumn{2}{@{}l}{\emph{Formats and rounding}}\\
$\mathrm{fl}_{16}(\cdot),\ \mathrm{fl}_{32}(\cdot)$ & Rounding of a real number to FP16 / FP32, round-to-nearest-even \\
$u_{16}=2^{-11}$ & FP16 unit roundoff (11-bit significand: 10 stored $+$ 1 implicit) \\
$u_{32}=2^{-24}$ & FP32 unit roundoff (24-bit significand: 23 stored $+$ 1 implicit) \\
RN, RZ & Round-to-nearest-even; round-toward-zero \\
$k$ &
\shortstack[l]{Inner-product length, i.e., the contraction dimension\\
in $A\in\mathbb{R}^{m\times k}$ and $B\in\mathbb{R}^{k\times n}$} \\
$\gamma_k=\dfrac{k u_{32}}{1-k u_{32}}$
& Worst-case rounding-error coefficient for accumulating $k$ terms in FP32 \\
$a=1+u_{16}$ & Shorthand used in the conservative componentwise bound \\
$\tau=u_{16}^2(3+2u_{16}+2u_{16}^2)$ & Exact-arithmetic truncation coefficient in Eq.~\ref{eq:truncation} \\
\midrule
\multicolumn{2}{@{}l}{\emph{Scalar decomposition (Sec.~\ref{method}--\ref{sec:accuracy_analysis})}}\\
$V$ & An FP32 scalar to be split \\
$S,E,M$ & Sign, biased exponent, and 23-bit stored mantissa of $V$ (Eq.~\ref{eq:fp32}) \\
$E_\text{offset}$ & Unbiased (offset) exponent of $V$, i.e. $E-127$ \\
$e_{\min},e_{\max}$ &
\shortstack[l]{Minimum and maximum unbiased exponents\\
of the finite FP16 high components} \\
$V_\text{high},V_\text{low}$ & FP16 high component and FP16 residual component of $V$ \\
$S',E'$ & Sign and biased 5-bit exponent of the FP16 components \\
$M_\text{high},M_\text{low}$ & Stored mantissas of $V_\text{high}$, $V_\text{low}$ \\
$c\in\{0,1\}$ & Carry produced when $V$ is rounded to $V_\text{high}$; $E-127+c=E'-15$ \\
$N$ &
\shortstack[l]{Residual-class index: $N=-1$ denotes the midpoint tie;\\
otherwise $0\le N\le 12$ counts leading zeros below the split} \\
$l_M=23$ & Number of stored mantissa bits of FP32 \\
$l_{M_\text{high}}=10$ & Number of stored mantissa bits of FP16 \\
$b_\text{low}=15$ & FP16 exponent bias \\
$s_f=2^{s_b}$ &
\shortstack[l]{Residual scaling factor; $s_b$ is the scaling exponent\\
($s_b=12$ in the evaluated exponent window)} \\
\midrule
\multicolumn{2}{@{}l}{\emph{Matrix level (Sec.~\ref{sec:err_bound}--\ref{sec:accum_order})}}\\
$A,B,C$ & FP32 input matrices ($m\times k$, $k\times n$) and output ($m\times n$) \\
$A_\text{half},B_\text{half}$ & FP16 high components of $A,B$; written $A^{[g]},B^{[g]}$ per $k$-block \\
$R_{A,\text{half}},R_{B,\text{half}}$ & Scaled FP16 residuals of $A,B$; written $R_A^{[g]},R_B^{[g]}$ per $k$-block \\
$g,G$ & Index and last index of the $k$-dimension blocks, $g=0,\dots,G$ \\
$|A|$ & Matrix of entrywise absolute values, $(|A|)_{ij}=|A_{ij}|$ \\
$\le$ (on matrices) & Entrywise inequality \\
\midrule
\multicolumn{2}{@{}l}{\emph{Blocking and performance (Sec.~\ref{system_optimizations}--\ref{performance})}}\\
$b_m,b_k,b_n$ & Block dimensions of the L1 tiles of $A$ and $B$ \\
$N_\text{fused}$ & Number of $A$ blocks resident in L1 simultaneously \\
$f$ & L1 fusion efficiency factor, $0.92\le f\le 1$ in our experiments \\
$N_\text{core}=32$ & Number of AI cores on Ascend 910A \\
$\mathrm{OI}_{\mathrm{proxy}}$ & Normalized intensity proxy used for the first-level performance diagnostic \\
\bottomrule
\end{tabular}
\end{table}

\subsection{FP32-to-FP16 Two-Component Representation}

According to the IEEE-754 floating-point standard~\cite{kahan1996ieee}, a single-precision (FP32) number is represented using 32 bits: 1 sign bit ($S$), 8 exponent bits ($E$), and 23 mantissa (fraction) bits ($M$), with an implicit leading bit for normalized values. The actual value $V$ is given by Eq.~\ref{eq:fp32}:
\begin{equation}
V = (-1)^S \times 2^{(E - 127)} \times (1.M)
\label{eq:fp32}
\end{equation}
Here, $E - bias$ is the offset exponent, and $M$ is a 23-bit fraction. For normalized numbers, the leading bit before the decimal point is implicitly assumed to be 1.

\begin{figure}[!t]
\centering
\includegraphics[width=3.6in]{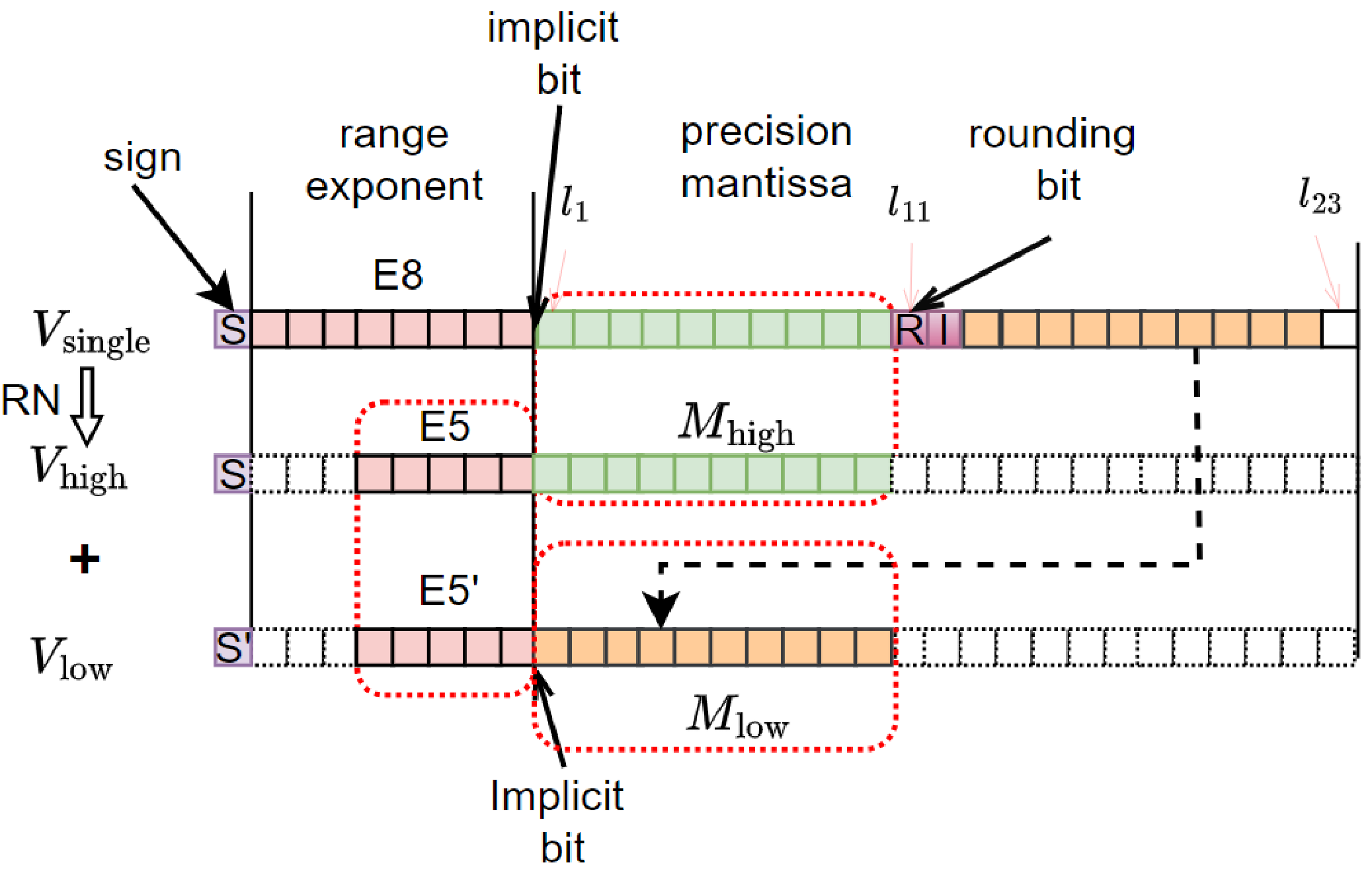}
\caption{Splitting a single FP32 floating number into two FP16 floating numbers}
\Description{Illustration of decomposing one FP32 value into a high FP16 component and a low residual FP16 component for precision recovery.}
\label{SP_HP}
\end{figure}

A half-precision (FP16) number uses 16 bits: 1 sign bit, 5 exponent bits, and 10 mantissa (fraction) bits, also with an implicit leading bit. To approximate FP32 precision using FP16 units, an FP32 value $V$ is decomposed into a high-order part ($V_{\text{high}}$) and a low-order residual part ($V_{\text{low}}$), as conceptually illustrated in Fig.~\ref{SP_HP}. This decomposition follows the round-to-nearest-even (RN) rule, which is critical for minimizing rounding errors.

To improve readability, we express $V_{\text{high}}$ and $V_{\text{low}}$ separately in Eq.~\ref{fp16_high_low}:

\begin{equation}
\begin{cases}
\begin{aligned}
V_{\text{high}} &= 
\begin{cases}
(-1)^S \times 2^{(E'-15)} \times 1.M_{\text{high}} 
& \text{if } 1 \leq E' \leq 30 \text{ (normalized)} \\
(-1)^S \times 2^{-14} \times 0.M_{\text{high}} 
& \text{if } E' = 0 \text{ (subnormal)}
\end{cases} \\
V_{\text{low}} &= 
\begin{cases}
\begin{array}[c]{@{}l@{}}
\hspace{1pt}(-1)^{S'} \times 2^{(E'-15-11-1-N)} \times 1.M_{\text{low}}
\end{array}
& \begin{array}[c]{@{}l@{}}
\text{if } 1 \leq (E'-11-1-N) \leq 30 \\
\text{(normalized)}
\end{array} \\
\begin{array}[c]{@{}l@{}}
\hspace{1pt}(-1)^{S'} \times 2^{-14} \times 0.M_{\text{low}}
\end{array}
& \begin{array}[c]{@{}l@{}}
\text{if } (E'-11-1-N) = 0 \\
\text{(subnormal)}
\end{array}
\end{cases}
\end{aligned}
\end{cases}
\label{fp16_high_low}
\end{equation}
Here, $E'$ is the biased 5-bit FP16 exponent, so that $E'-15$ and $E'-12-N-15$ are the offset exponents of the high and low parts, and $M_{\text{high}}$, $M_{\text{low}}$ are their stored mantissas. The symbol $c$ denotes the carry introduced by rounding the high part, $E - 127 + c = E' - 15$, with $c = 1$ if the rounding carries and $c=0$ otherwise. When $c = 1$, the residual $V-V_\text{high}$ is negative and the low part's sign bit differs from $S$; this sign flip is a direct consequence of using RN rather than RZ for the high-part conversion and was first pointed out by Ootomo and Yokota~\cite{ootomo2022recovering}. Finally, $N$ indexes the residual classes: $N=-1$ denotes the midpoint tie used in the probability model below, while ordinary residuals have $0\leq N\leq l_M-l_{M_\text{high}}-1=12$.

The original FP32 value is then reconstructed as $V \approx V_{\text{high}} + V_{\text{low}}$. Since RN gives $|V-V_\text{high}|\le u_{16}|V|$ and the second conversion incurs a further relative error of at most $u_{16}$, the two-word representation satisfies
\begin{equation}
\bigl|V - (V_\text{high} + V_\text{low})\bigr| \;\le\; u_{16}^2\,|V| \;=\; 2^{-22}\,|V|,
\label{eq:split_error}
\end{equation}
provided $V_\text{high}$ remains a normal FP16 value (or zero) and the scaled residual is represented without underflow or overflow. Under these conditions, the pair $(V_\text{high},V_\text{low})$ carries $11+11=22$ significand bits, two fewer than the 24 of FP32 and eleven
more than a standalone FP16 value, and behaves like a format of unit roundoff $2^{-22}$. Note that the residual $V - V_\text{high}$ is itself exact in FP32, since it needs at most $l_M+1=24$ significand bits; multiplication by the power of two $s_f$ is exact as well. Nothing is lost before the second rounding; underflow or overflow in either FP16 component is what can invalidate Eq.~\ref{eq:split_error}. Sec.~\ref{sec:accuracy_analysis} takes up that question. The cost of all this is extra conversion work and storage, in return for eleven additional bits on hardware that offers FP16 matrix engines and nothing better.

\section{RN-Based Accuracy and Range Analysis}
\label{sec:accuracy_analysis}

Equation~\ref{eq:split_error} holds only while neither FP16 component underflows or overflows. This section establishes when those assumptions fail and how the residual scaling factor must be bounded to keep the representation valid. Mary and Mikaitis~\cite{mary2025narrow} treat the general version of the question, in which it is the narrow exponent range of a low-precision format, not its short significand, that limits the accuracy of a matrix product. What follows is the FP16 instance, worked out for RN conversion. Several earlier implementation papers assumed RZ instead; the difference moves the boundary cases and accounts for the scaling factor adopted here being different from theirs.

\subsection{Underflow and Gradual Underflow under RN}

Underflow and gradual underflow primarily affect the low-order component $V_{\text{low}}$, because of its smaller magnitude. Throughout this subsection, \emph{gradual underflow} means that the magnitude of the exact residual lies below the smallest normal FP16 value, $2^{-14}$, whereas \emph{underflow} means that it lies below the smallest positive FP16 subnormal, $2^{-24}$. These are range events for the exact residual and should not be interpreted as saying that round-to-nearest conversion necessarily produces zero. In particular, a value below $2^{-24}$ but sufficiently close to it may round to the smallest FP16 subnormal, whereas
values below the halfway threshold $2^{-25}$ round to zero, with the exact midpoint handled by the ties-to-even rule.

Ascend converts with RN, so we compute the probabilities of these range events under that assumption. For an FP32 input whose mantissa bits below the split point are uniformly distributed, let $P(X, N=n)$ denote the joint probability that the residual has $n$ leading zeros \emph{and} that the high-part conversion is of type $X$, where $X=T$ (the high part is truncated, $c=0$) or $X=R$ (the high part is rounded up, $c=1$):

\begin{equation}
P(X, N=n) = \begin{cases}
0 & \text{if } n < -1 \\
\left(\frac{1}{2}\right)^{l_{M}-l_{M_\text{high}}+1} & \text{if } n = -1 \text{ and } X = T \text{ or } R \\
\left(\frac{1}{2}\right)^{n+2} & \text{if } 0 \leq n < l_{M}-l_{M_\text{high}}-1 \text{ and } X = T \text{ or } R \\
\left(\frac{1}{2}\right)^{l_{M}-l_{M_\text{high}}} & \text{if } n = l_{M}-l_{M_\text{high}}-1 \text{ and } X = T \\
0 & \text{if } n = l_{M}-l_{M_\text{high}}-1 \text{ and } X = R
\end{cases}
\end{equation}
Here $l_M = 23$ and $l_{M_\text{high}} = 10$ are the numbers of stored mantissa bits of FP32 and FP16 (Table~\ref{tab:notation}). The special case $n = -1$ corresponds to the tie in which the 11th mantissa bit is 1 and all remaining bits are 0. The cases above are exhaustive and the probabilities sum to one.

The first condition identifies residuals whose exact magnitudes enter the FP16 subnormal range, while the second identifies residuals whose exact magnitudes fall below the smallest positive FP16 subnormal:

\begin{equation}
\begin{cases}
E_\text{offset}-l_{M_\text{high}}+b_\text{low}-3 < N & \text{(gradual underflow)}\\
E_\text{offset}+b_\text{low}-3 < N & \text{(underflow)}
\end{cases}
\end{equation}
where $E_\text{offset}=E-127$ is the offset exponent of the FP32 input and $b_\text{low}=15$ is the FP16 exponent bias. Summing $P(X,N=n)$ over the values of $N$ that trigger each condition gives

\begin{equation}
\begin{cases}
P_{u+gu}(E_\text{offset})=\sum\limits_{N=E_\text{offset}-l_{M_\text{high}}+b_\text{low}-2}^{l_{M}-l_{M_\text{high}}-1}(P(T, N)+P(R, N)) \\
P_{u}(E_\text{offset})=\sum\limits_{N=E_\text{offset}+b_\text{low}-2}^{l_{M}-l_{M_\text{high}}-1}(P(T, N)+P(R, N))
\end{cases}
\end{equation}

Accordingly, $P_{u+gu}$ is the probability that the exact residual lies in or below the FP16 subnormal range, whereas $P_u$ is the probability that it lies below $2^{-24}$. The latter is not the probability that the RN conversion returns zero.

As shown in Fig.~\ref{underflow_precision_bits}(a), residual-range loss becomes a significant concern when the offset exponent of the original FP32 number is small. If subnormals are not supported by the hardware, the probability that the exact residual enters the subnormal range exceeds 10\% at $E_\text{offset}=0$. If subnormals are supported, the probability that the exact residual falls below the smallest positive FP16 subnormal becomes significant only below $E_\text{offset}=-10$ and approaches 100\% for $E_\text{offset}<-12$. This statement concerns the magnitude
of the exact residual, not the probability that RN conversion produces zero.

\subsection{Residual Scaling Strategy}
\label{scaling}

In practice the value of $N$, the number of leading zeros of
$V_{\text{low}}$, is not known to the caller. As the exponent of the exact residual drops below $-14$, the residual enters the FP16 subnormal range and progressively loses effective significand bits. If its exponent drops below $-24$, its magnitude is smaller than the smallest positive FP16 subnormal. This does not imply that every such residual is converted to zero under RN: values sufficiently close to $2^{-24}$ may round to the smallest subnormal, whereas sufficiently smaller values round to zero.
In either case, the 22-bit representation guarantee no longer holds. Scaling the low part before conversion is therefore essential.

\vspace{-\baselineskip}
\begin{center}
\fbox{\parbox{\linewidth}{
\textbf{Rule 1 (Underflow Mitigation):} Under RN rounding, when the absolute value of an FP32 number is below $2^{-2}$, the low part must be scaled to preserve 22 bits. If the value is below $2^{-12}$, the exact residual can fall below the smallest positive FP16 subnormal, so decomposition without sufficient scaling no longer guarantees preservation of the low component.
}}
\end{center}

It is important to note that once $V_\text{high}$ enters the FP16 subnormal range, which can occur for inputs below the normal-FP16 threshold $2^{-14}$, scaling only $V_\text{low}$ does not preserve the 22-bit relative-error guarantee. Whole-input scaling or explicit exponent
management for the high component is then required, which is beyond the current scope of this work.

Conversely, amplifying $V_{\text{low}}$ introduces the risk of overflow if the scaled residual exceeds the representable upper range of FP16. Let $e_{\min}=(E'-15)_{\min}$ and $e_{\max}=(E'-15)_{\max}$ denote the smallest and largest unbiased exponents of the finite FP16 high components in the input. Under RN, a sufficient upper bound follows from $|V-V_{\text{high}}|\leq 2^{e_{\max}-11}$: the scaled residual remains in range if $e_{\max}-11+s_b\leq 15$. This gives the following rule.

\vspace{-\baselineskip}
\begin{center}
\fbox{\parbox{\linewidth}{
\textbf{Rule 2 (Overflow Prevention):} Choose $s_b\leq 26-e_{\max}$. In particular, $s_b=12$ is guaranteed safe only when $e_{\max}\leq 14$, unless residual overflow is explicitly detected and handled.
}}
\end{center}

Rules 1 and 2 give sufficient lower and upper bounds on the scaling exponent. The lower bound allows residuals below the target $u_{16}^2|V|$ accuracy level to round away, while the upper bound prevents the largest scaled residual from overflowing:

\begin{equation}
-3 - (E'-15)_{\text{min}} \leq s_b \leq 26 - (E'-15)_{\text{max}}.
\label{sf_bounds}
\end{equation}

The choice of $s_b$ therefore depends on the input exponent window. If the finite high components span the entire normal FP16 range, with $(E'-15)_{\text{min}}=-14$ and $(E'-15)_{\text{max}}=15$, Eq.~\ref{sf_bounds} gives the unique fixed choice $s_b=11$. The value $s_b=12$ extends the protected low-magnitude range by one exponent but requires $(E'-15)_{\text{max}}\leq 14$ or an overflow guard. For example, $V=32784$ rounds to $V_{\text{high}}=32768$, leaving a residual of $16$; multiplying that residual by $2^{12}$ gives $65536$, which overflows FP16. The experiments below use $s_b=12$ for the evaluated low-magnitude exponent window rather than as a guarantee over the full FP16 range.

\begin{figure}[!t]
    \centering
    \begin{subfigure}{0.49\textwidth}
        \centering
        \includegraphics[width=0.9\linewidth]{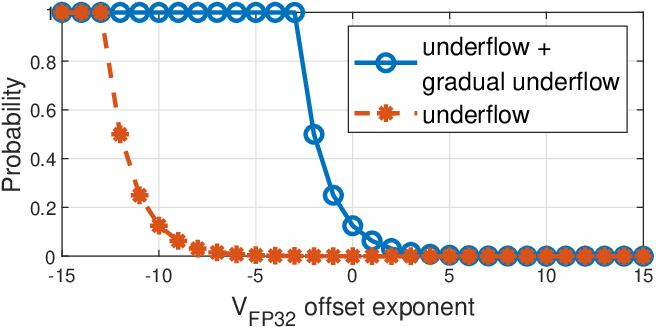}
        \subcaption{Probability of subnormal-range and below-subnormal-range residuals}
    \end{subfigure}
    \begin{subfigure}{0.49\textwidth}
        \centering
        \includegraphics[width=0.9\linewidth]{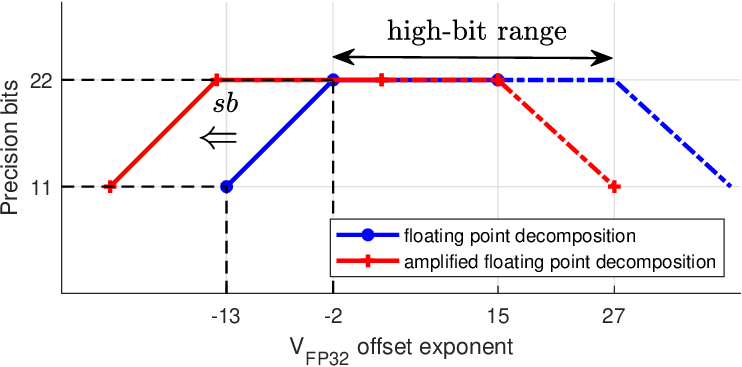}
        \subcaption{Precision bits and FP32 exponent}
    \end{subfigure}
    \caption{Analysis of residual range loss and retained precision bits}
    \Description{Two plots showing the numerical behavior of FP32-to-FP16 decomposition under round-to-nearest conversion. The first plot shows the probabilities that the exact residual enters the FP16 subnormal range or falls below the smallest positive FP16 subnormal. The second plot shows retained precision bits with and without residual scaling across different exponent ranges.}
    \label{underflow_precision_bits}
\end{figure}

Based on Eq.~\ref{sf_bounds}, Fig.~\ref{underflow_precision_bits}(b) shows how the number of retained significand bits varies with the offset exponent of the original FP32 input, both with and without scaling. Without amplification, precision degrades gradually as the exponent decreases, particularly for values approaching the FP16 subnormal range. Setting $s_b=12$ shifts the curve left by 12 exponents and widens the high-precision region toward small inputs. Without scaling, the exact residual sits about 12 exponents below the input, so it enters the FP16
subnormal range once $E_\text{offset}<-2$ and falls below the smallest positive FP16 subnormal once $E_\text{offset}<-12$. These are range boundaries for the exact residual, rather than exact zero-rounding thresholds. Multiplying the residual by $2^{12}$ before conversion moves the two range boundaries to $E_\text{offset}<-14$ and $E_\text{offset}<-24$, respectively. The corresponding upper-range condition is $e_{\max}\leq14$, as stated in Rule 2.

Both rules instantiate a general principle for narrow-range formats: scale the residual as far from the underflow threshold as the largest entry permits, so that the remaining underflow error stays below the rounding-error target without causing overflow. Mary and Mikaitis~\cite{mary2025narrow} state and analyze this principle for arbitrary formats. Equation~\ref{sf_bounds} is its sufficient FP16/RN specialization for the two-word representation. The experimental choice $s_b=12$ is therefore range-dependent; $s_b=11$ is the conservative fixed choice over the full normal-FP16 high-component range.

\subsection{Three-Term Reconstruction and the Omitted Low-Low Product}
\label{sec:three_term}

Converting FP32 inputs directly to FP16 reduces the significand to 11 bits. SGEMM-cube instead applies the two-word split of Sec.~\ref{method} to both operands and expands the product, as formulated in Eq.~\ref{recover_eqn}. Although the dominant multiplications are executed by FP16 Cube units, the reconstruction and accumulation of the terms are performed in FP32: Ascend NPUs provide FP32 scalar/vector arithmetic, FP32 accumulation in the L0C buffer, and FP32 memory storage, but no high-throughput native FP32 matrix engine. The method therefore uses FP16 matrix units for the throughput-critical operations but is not an FP16-only arithmetic pipeline.

\begin{equation}
\left\{
\begin{aligned}
A_{\text{half}} &= \text{to\_half}(A_{\text{single}}), \quad R_{A,\text{half}} = \text{to\_half}((A_{\text{single}} - \text{to\_single}(A_{\text{half}})) \times s_f) \\
B_{\text{half}} &= \text{to\_half}(B_{\text{single}}), \quad R_{B,\text{half}} = \text{to\_half}((B_{\text{single}} - \text{to\_single}(B_{\text{half}})) \times s_f) \\
C_{\text{single}} &= A_{\text{single}} B_{\text{single}} \\
&\approx (A_{\text{half}} + R_{A,\text{half}} / s_f)(B_{\text{half}} + R_{B,\text{half}} / s_f) \\
&= \underbrace{A_{\text{half}} B_{\text{half}} + R_{A,\text{half}} B_{\text{half}} / s_f + A_{\text{half}} R_{B,\text{half}} / s_f}_{\text{computed: three FP16 GEMMs}}
 \;+\; \underbrace{R_{A,\text{half}} R_{B,\text{half}} / {s_f}^2}_{\text{dropped, bounded by Eq.~\ref{eq:ll_bound}}}
\end{aligned}
\right.
\label{recover_eqn}
\end{equation}

Here $A_{\text{single}}, B_{\text{single}}$ are the FP32 inputs and $C_{\text{single}}$ the FP32 output, while the high parts $A_{\text{half}}, B_{\text{half}}$ and the scaled residuals $R_{A,\text{half}}, R_{B,\text{half}}$ are FP16. The splitting formula, the amplification of the residual by $s_f$ and the decision to drop the low-low product all come from Ootomo and Yokota~\cite{ootomo2022recovering}. They are restated here because the constants derived in Sec.~\ref{sec:err_bound} depend on them. For the remainder of this section we abbreviate $A \equiv A_{\text{single}}$ and $B \equiv B_{\text{single}}$.

The low-low term is not zero, and calling it negligible without a bound settles nothing. By RN the unrounded residual of $A$ satisfies $|A - A_{\text{half}}| \le u_{16}|A|$, and the subsequent FP16 conversion inflates it by at most a further factor $(1+u_{16})$, giving $|R_{A,\text{half}}/s_f| \le u_{16}(1+u_{16})|A|$ entrywise, and similarly for $B$. The dropped term then satisfies
\begin{equation}
\Bigl| R_{A,\text{half}} R_{B,\text{half}} / s_f^2 \Bigr| \;\le\; u_{16}^2(1+u_{16})^2\,|A|\,|B| \;=\; 2^{-22}\,|A|\,|B| \;+\; O(u_{16}^3),
\label{eq:ll_bound}
\end{equation}
where $|\cdot|$ and $\le$ act entrywise and $|A|\,|B|$ is the ordinary matrix product of the nonnegative matrices $|A|$ and $|B|$. The leading coefficient is $u_{16}^2=2^{-22}=4u_{32}$. Retaining the factor $(1+u_{16})^2$, the exact coefficient is slightly larger than $\gamma_4$ but smaller than $\gamma_5$; the omitted term therefore has the same leading-order scale as a length-four FP32 accumulation bound and is dominated by the classical FP32 accumulation bound for $k\geq5$.

Three error sources are left: the two-word representation error of Eq.~\ref{eq:split_error}, the FP32 accumulation inside each of the three GEMMs, and the final combination of the three result matrices. RN matters for all three, since RZ would introduce a systematic bias that does not cancel. We assume throughout that the Cube unit accumulates its FP16 products into FP32 with RN. Under that assumption Ascend differs from NVIDIA Tensor Cores, whose internal accumulation rounds toward zero~\cite{fasi2021numerical}, the effect to which Fasi et al.~\cite{fasi2023multiword} attribute the observed accuracy gap between double-fp16 and fp32 on GPUs. SGEMM-cube would then have no need for the ``accumulate outside the matrix unit'' correction that Ootomo and Yokota~\cite{ootomo2022recovering} introduced on the A100. We have not verified the Cube's internal rounding directly, and a characterization in the style of~\cite{fasi2021numerical} would settle the point; see Sec.~\ref{sec:limitations}.

\subsection{A Componentwise Error Bound}
\label{sec:err_bound}

We now bound the error of the complete scheme. The result is the two-word FP16/FP32 specialization of the multiword analysis of Fasi et al.~\cite{fasi2023multiword}, stated with the constants written out, since it is the constants that decide whether the scheme is usable at a given $k$ and a given accuracy target.

Write $\tilde R_A = R_{A,\text{half}}/s_f$ for the descaled residual and let $E_A = A - A_{\text{half}} - \tilde R_A$ be the representation error, so that by Eq.~\ref{eq:split_error}
\begin{equation}
|E_A| \le u_{16}^2 |A|, \qquad |\tilde R_A| \le u_{16}(1+u_{16})|A|, \qquad |A_{\text{half}}| \le (1+u_{16})|A|,
\label{eq:split_ineqs}
\end{equation}
entrywise, and analogously for $B$. Let $a=1+u_{16}$. Expanding $AB=(A_{\text{half}}+\tilde R_A+E_A)(B_{\text{half}}+\tilde R_B+E_B)$ and subtracting the three terms actually computed leaves the exact-arithmetic truncation
\begin{equation}
T = \tilde R_A \tilde R_B + E_A B + (A_{\text{half}}+\tilde R_A)E_B,
\qquad
|T| \le \tau\,|A||B|,
\label{eq:truncation}
\end{equation}
where we used $|A_{\text{half}}+\tilde R_A|=|A-E_A|\leq(1+u_{16}^2)|A|$, and
\begin{equation}
\tau = u_{16}^2\bigl(3+2u_{16}+2u_{16}^2\bigr).
\end{equation}
We now analyze the termwise reconstruction, in which the three GEMM result matrices are accumulated separately and combined by two final FP32 additions. A product of two FP16 numbers is exactly representable in FP32, so the three length-$k$ FP32 accumulations contribute at most
$\gamma_k a^2(1+2u_{16})|A||B|$. The two final additions contribute at most
$\gamma_2(1+\gamma_k)a^2(1+2u_{16})|A||B|$. Hence the computed termwise result $\widehat C$ satisfies the conservative componentwise bound
\begin{equation}
\bigl|AB - \widehat{C}\bigr|
\;\le\;
\Bigl[
\tau
+ \gamma_k a^2(1+2u_{16})
+ \gamma_2(1+\gamma_k)a^2(1+2u_{16})
\Bigr]|A||B|.
\label{eq:main_bound}
\end{equation}
Expanding this coefficient to first order gives
\begin{equation}
\bigl|AB - \widehat{C}\bigr|
\;\le\;
\bigl(\gamma_k+14u_{32}\bigr)|A||B|
+O\!\left(u_{16}^3+\gamma_k u_{16}\right)|A||B|.
\label{eq:main_bound_first_order}
\end{equation}
Thus $\gamma_k+14u_{32}$ is the first-order form of the bound, not the exact coefficient after the higher-order terms are discarded.

Several consequences follow.

The classical bound for an FP32 GEMM is $|A B - \widehat C_{32}| \le \gamma_k |A||B|$. Equation~\ref{eq:main_bound_first_order} shows that the leading scheme-specific contribution is the additive constant $14u_{32}$, since $u_{16}^2=2^{-22}=4u_{32}$. The exact coefficient retained in Eq.~\ref{eq:main_bound} is slightly larger than this first-order expression; its ratio to the classical FP32 bound falls below $2$ at $k=15$ and tends to $1$ as $k$ grows. Past a contraction depth of a few tens, FP32 accumulation is therefore the dominant term in the bound. This agrees with Fig.~\ref{mkn_error}, where the measured error is essentially independent of $m$ and $n$ but varies with the accumulation depth $k$.

The dropped low-low term is not the dominant contribution for practical contraction depths. Its exact coefficient is $u_{16}^2(1+u_{16})^2$; this is slightly larger than $\gamma_4$ but smaller than $\gamma_5$, so the classical FP32 accumulation coefficient exceeds it for $k\geq5$.

Since the bound is componentwise it is also a max-norm bound. Dividing Eq.~\ref{eq:main_bound} entrywise, the relative error of entry $(i,j)$ is bounded by the coefficient in Eq.~\ref{eq:main_bound} multiplied by $\kappa_{ij}=(|A||B|)_{ij}/|(AB)_{ij}|$, the usual cancellation measure. This quantity equals one for nonnegative data and can be large when an inner product cancels. The bound therefore explains why the symmetric-sampling results in Fig.~\ref{scaling_bit_error}(a) can have larger relative errors than the non-negative results in Fig.~\ref{scaling_bit_error}(b). An empirical max-norm study remains to be done and is listed in Sec.~\ref{sec:limitations}.

Eq.~\ref{eq:main_bound} assumes finite intermediate values, no underflow or overflow in either FP16 component, exact representation of FP16 products in FP32, and RN FP32 accumulation. Outside the exponent window of Sec.~\ref{scaling}, the effective representation error can degrade toward native FP16 accuracy, as visible at small offset exponents in Fig.~\ref{underflow_precision_bits}(b).

\subsection{Elementwise vs. Termwise Accumulation}
\label{sec:accum_order}

\begin{figure}[!t]
    \centering
    \begin{subfigure}{0.49\textwidth}
        \centering
        \includegraphics[width=0.9\linewidth]{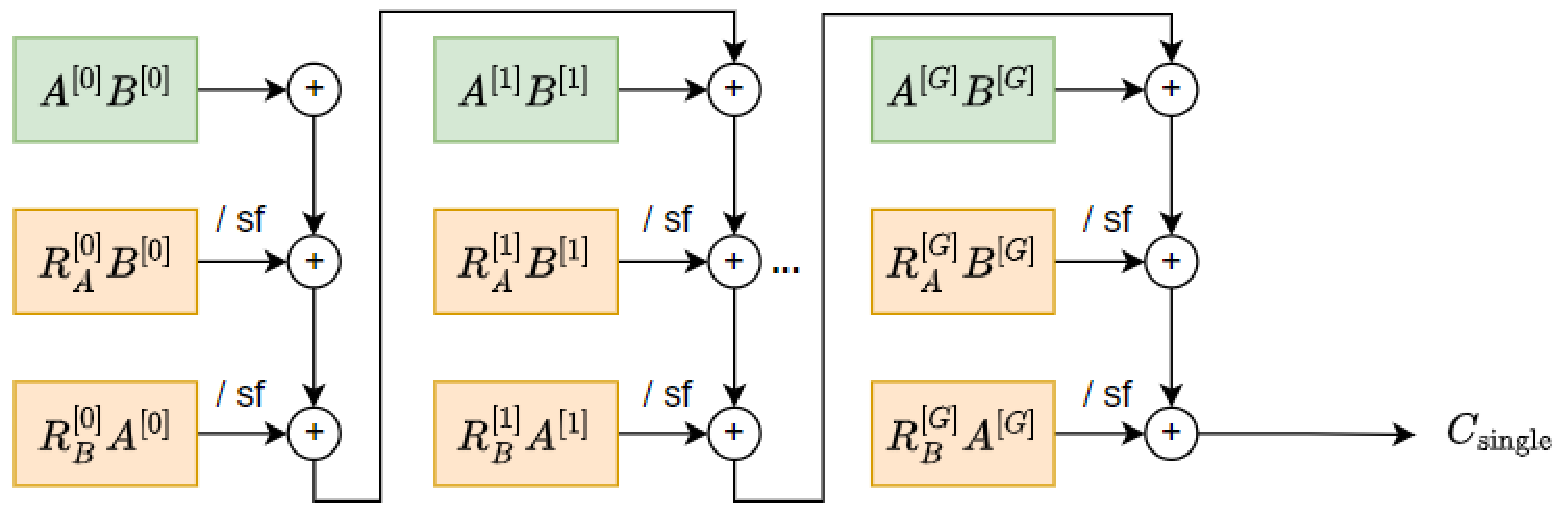}
        \caption{elementwise computation}
    \end{subfigure}
    \begin{subfigure}{0.49\textwidth}
        \centering
        \includegraphics[width=0.9\linewidth]{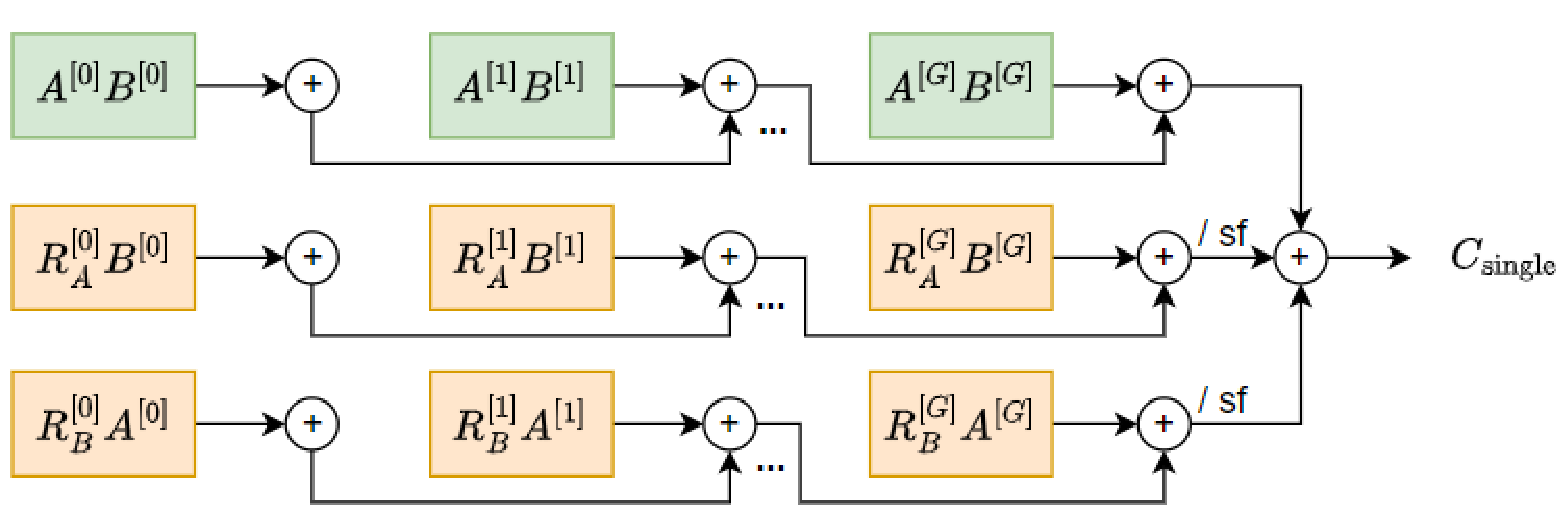}
        \caption{termwise computation}
    \end{subfigure}
    \caption{Computation sequences of the SGEMM-cube precision recovery algorithm. Superscript $[g]$ indexes the block of the contraction dimension, $g=0,\dots,G$; green boxes are $A^{[g]}B^{[g]}$, orange boxes are $R_A^{[g]}B^{[g]}$ and $A^{[g]}R_B^{[g]}$, and ``$+$'' nodes are FP32 additions. In (a) the three terms of each block are folded into a single running accumulator; in (b) each of the three term streams is accumulated separately and the three results are combined once at the end.}
    \Description{Flowcharts comparing two accumulation strategies for SGEMM-cube. The left diagram illustrates elementwise accumulation of the decomposition terms, where the three products of each k-block are added into one running accumulator. The right diagram illustrates termwise accumulation where the high-high products and the two correction matrices are accumulated in three separate streams before a single final reconstruction.}
    \label{flowchart}
\end{figure}

Equation~\ref{eq:main_bound} analyzes the termwise organization: three separately accumulated GEMM result matrices followed by two FP32 additions. The exact-arithmetic truncation is common to both organizations, but the elementwise implementation performs a different number and ordering of block-level additions and therefore requires a separate rounding analysis. We compare the two organizations empirically below, and Fig.~\ref{flowchart} shows both.

The notation in the two diagrams is as follows. The contraction dimension is partitioned into blocks indexed by $g = 0, 1, \dots, G$, one block per iteration of the $k$-loop of Algorithm~\ref{alg:double_buffer}. For block $g$, $A^{[g]}$ and $B^{[g]}$ are the FP16 high components of the corresponding sub-blocks of $A_{\text{half}}$ and $B_{\text{half}}$, and $R_A^{[g]}$, $R_B^{[g]}$ the matching scaled FP16 residual blocks from $R_{A,\text{half}}$ and $R_{B,\text{half}}$. Each shaded box is one FP16 Cube GEMM over one $k$-block: green for the high-high products $A^{[g]}B^{[g]}$, orange for the two cross terms $R_A^{[g]}B^{[g]}$ and $A^{[g]}R_B^{[g]}$, which have to be divided by $s_f$ before they can join the high-high term. Every ``$+$'' node is an FP32 addition. Written out, Eq.~\ref{recover_eqn} becomes $C_{\text{single}} \approx \sum_{g=0}^{G}\bigl(A^{[g]}B^{[g]} + (R_A^{[g]}B^{[g]} + A^{[g]}R_B^{[g]})/s_f\bigr)$, and the two strategies place the parentheses differently.

\paragraph{Elementwise accumulation (Fig.~\ref{flowchart}(a))} For each block $g$ in turn, the three products of that block are reduced into a single running FP32 accumulator, which is then carried into block $g+1$:
\begin{equation}
\widehat C \;=\; \Bigl(\!\cdots\Bigl(\bigl(A^{[0]}B^{[0]} + R_A^{[0]}B^{[0]}/s_f\bigr) + A^{[0]}R_B^{[0]}/s_f\Bigr) + A^{[1]}B^{[1]} + \cdots\Bigr).
\label{eq:elementwise}
\end{equation}
Only one accumulator is needed, which makes this the cheaper option in L0C and UB occupancy. The weakness is that every correction term of magnitude $O(u_{16})$ goes directly into a running sum of magnitude $O(1)$, so each addition can lose the correction's low-order bits to swamping.

\paragraph{Termwise accumulation (Fig.~\ref{flowchart}(b))} The three term streams are accumulated separately over all blocks and combined once at the end:
\begin{equation}
\widehat C \;=\; \sum_{g=0}^{G} A^{[g]}B^{[g]} \;+\; \frac{1}{s_f}\Bigl(\sum_{g=0}^{G} R_A^{[g]}B^{[g]}\Bigr) \;+\; \frac{1}{s_f}\Bigl(\sum_{g=0}^{G} A^{[g]}R_B^{[g]}\Bigr).
\label{eq:termwise}
\end{equation}
Each of the three sums now aggregates terms of comparable magnitude. The corrections accumulate among themselves at full FP32 precision and meet the dominant product only once, in the final addition, rather than $G+1$ times. Two extra accumulators are the price, meaning more UB pressure and one additional pass of vector adds. We use this variant by default; Sec.~\ref{error} shows it to be the more accurate of the two in the small-exponent regime.

Eq.~\ref{eq:elementwise} is not the best elementwise order available. Summing the two cross terms within a block before adding the high-high product,
$\bigl((R_A^{[g]}B^{[g]} + A^{[g]}R_B^{[g]})/s_f\bigr) + A^{[g]}B^{[g]}$,
lets the corrections combine at their own scale at no additional cost, and is a per-block version of what makes Eq.~\ref{eq:termwise} work. Fig.~\ref{flowchart}(a) reflects the order as we originally implemented it. Reordering this way should recover part of the gap to termwise accumulation but not all of it, since it removes the swamping inside a block and not across blocks. We have not measured how much.

\section{Implementation and Optimization on Ascend NPU}

\label{system_optimizations}

\subsection{Ascend 910A Architecture}
\label{sec:architecture}
\begin{figure}[!t]
\centering
\includegraphics[width=3.6in]{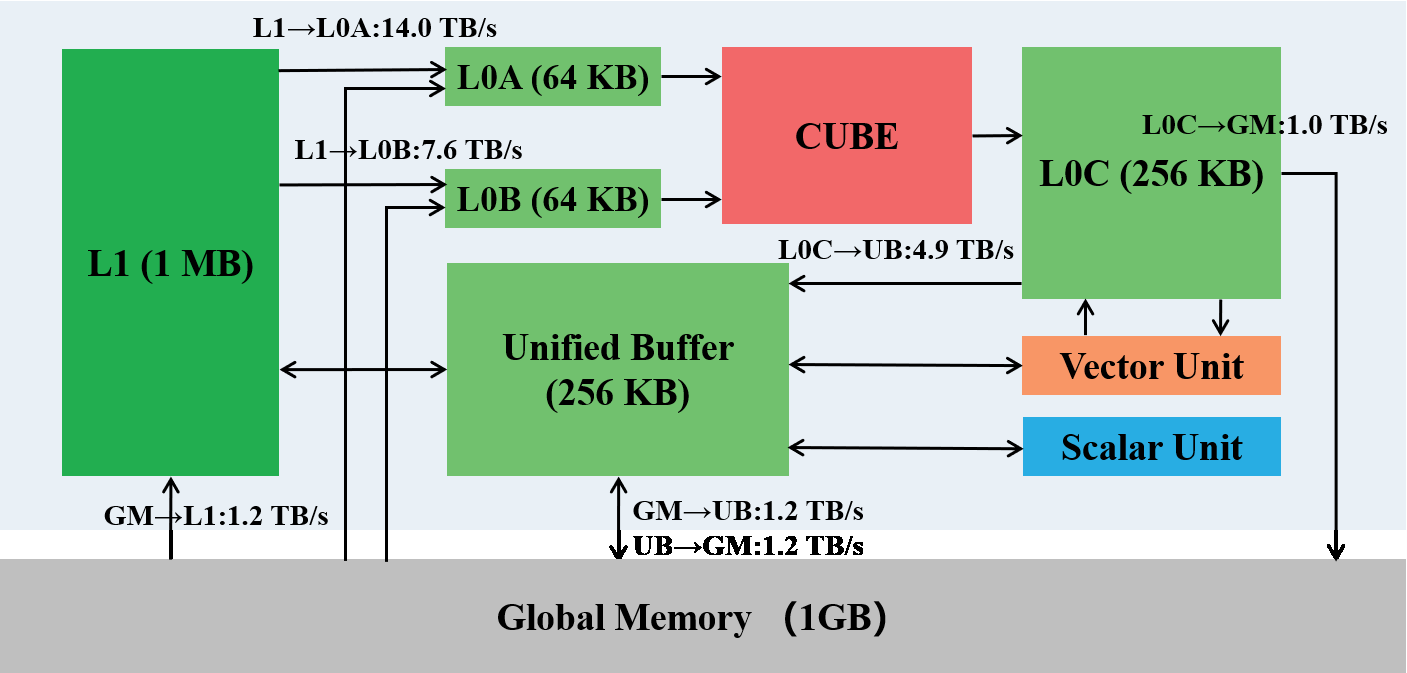}
\caption{DaVinci architecture of Huawei Ascend NPU AI core}
\Description{Architecture diagram of one Ascend AI core showing FP16 Cube units, L0 buffers, L1 cache, unified buffer, DMA engines, and memory hierarchy connections.}
\label{AI_core_architecture}
\end{figure}

The Huawei Ascend 910A NPU adopts the DaVinci architecture (Fig.~\ref{AI_core_architecture}), where each AI core integrates FP16 Cube units, a hierarchical memory system (L0A, L0B, L0C, L1, and a Unified Buffer, UB), and dedicated DMA engines. The aggregate FP16 Cube throughput is 256\,TFLOP/s across 32 AI cores; there are no native FP32 GEMM units. FP32 workloads must therefore be approximated on FP16-only engines, which raises challenges in both numerical accuracy and system efficiency.

Two properties of this hierarchy shape the implementation, and both can be put in numbers. The first is machine balance. Dividing 256\,TFLOP/s by 1.2\,TB/s of HBM bandwidth gives $\approx\!213$ FLOP/Byte, so a kernel has to sustain at least that much arithmetic per byte fetched to stay compute-bound. An NVIDIA A100 SXM (312\,TFLOP/s FP16 tensor, 2.04\,TB/s) sits at $\approx\!153$ FLOP/Byte and an H100 SXM (989\,TFLOP/s, 3.35\,TB/s) at $\approx\!295$ FLOP/Byte. The 910A is therefore about $1.4\times$ more compute-dense per byte of bandwidth than the A100, though not more so than the latest GPUs; its balance sits at the demanding end of the current range without being unique. Under the three-GEMM FP32-equivalent convention the corresponding operation-count ratio is $85.3/1.2 \approx 71$ FLOP/Byte of logical FP32-equivalent work; Sec.~\ref{performance} uses it only as part of a first-level diagnostic rather than as a complete bottleneck classification. The second property matters more. The hierarchy is managed entirely in software: every main-memory\,$\to$\,L1, L1\,$\to$\,L0A/L0B and L0C\,$\to$\,UB\,$\to$\,main-memory movement is issued explicitly by the kernel, with no hardware cache to absorb a poor blocking choice. This, more than the machine balance, is what stops GPU-tuned schedules from porting.

\subsection{Relation to Standard GEMM Engineering}

The blocking and double-buffering techniques described below are standard practice in high-performance GEMM and follow the structure laid down by Goto and van de Geijn~\cite{goto2008anatomy}. We claim no novelty for them. Two things in this section are specific to our setting. One is their instantiation on a memory hierarchy managed entirely in software, where $N_\text{fused}$ provides an L1-capacity constraint and a traffic proxy guides the initial tile search before measured tuning. The other is that the decomposition needs three GEMMs, not one, so the same double-buffered pipeline (Algorithm~\ref{alg:double_buffer}) is instantiated once per term of Eq.~\ref{recover_eqn}, and the L1 and L0 budgets of Eq.~\ref{block_limit} have to hold both the high and the residual FP16 operand copies. A reader already familiar with GEMM engineering can skip to the reuse proxy and the block-size constraints of Eq.~\ref{block_limit}.

The numerical formulation of Sec.~\ref{sec:accuracy_analysis} buys accuracy over direct FP16 GEMM, but throughput on Ascend has to be earned separately. The DaVinci memory hierarchy is the binding constraint. Bandwidth from main memory to L1 is low relative to Cube throughput, no hardware cache compensates for a poor tiling choice, and a naive pipeline leaves the Cube units idle. We address these with L1-aware blocking, which cuts memory traffic by maximizing reuse of the resident $A$ tile, and double-buffered pipelining, which overlaps transfers with computation.

\subsection{L1-Aware Blocking}
\label{sec:cache_blocking}

Efficient utilization of the L1 buffer is critical for sustaining high throughput on Ascend NPUs, where the bandwidth from main memory to L1 is substantially lower than that from L1 to the compute units (Fig.~\ref{AI_core_architecture}). Without careful blocking, frequent L1 refills stall the FP16 cube units, resulting in under-utilization of compute resources. 

To address this bottleneck, we develop an \emph{L1-aware blocking strategy} tailored to the hierarchical memory design of Ascend. The design is guided by two principles:

(1) Maximizing L1 Reuse for Matrix A: Blocks of matrix A are designed to reside in L1 for as long as possible, maximizing their reuse across multiple inner-loop iterations.

(2) Enabling Double Buffering for Matrix B: Space is reserved in L1 for double buffering of matrix B blocks, allowing for alternating access and prefetching.

\begin{figure}[!t]
\centering
\includegraphics[width=4.9in]{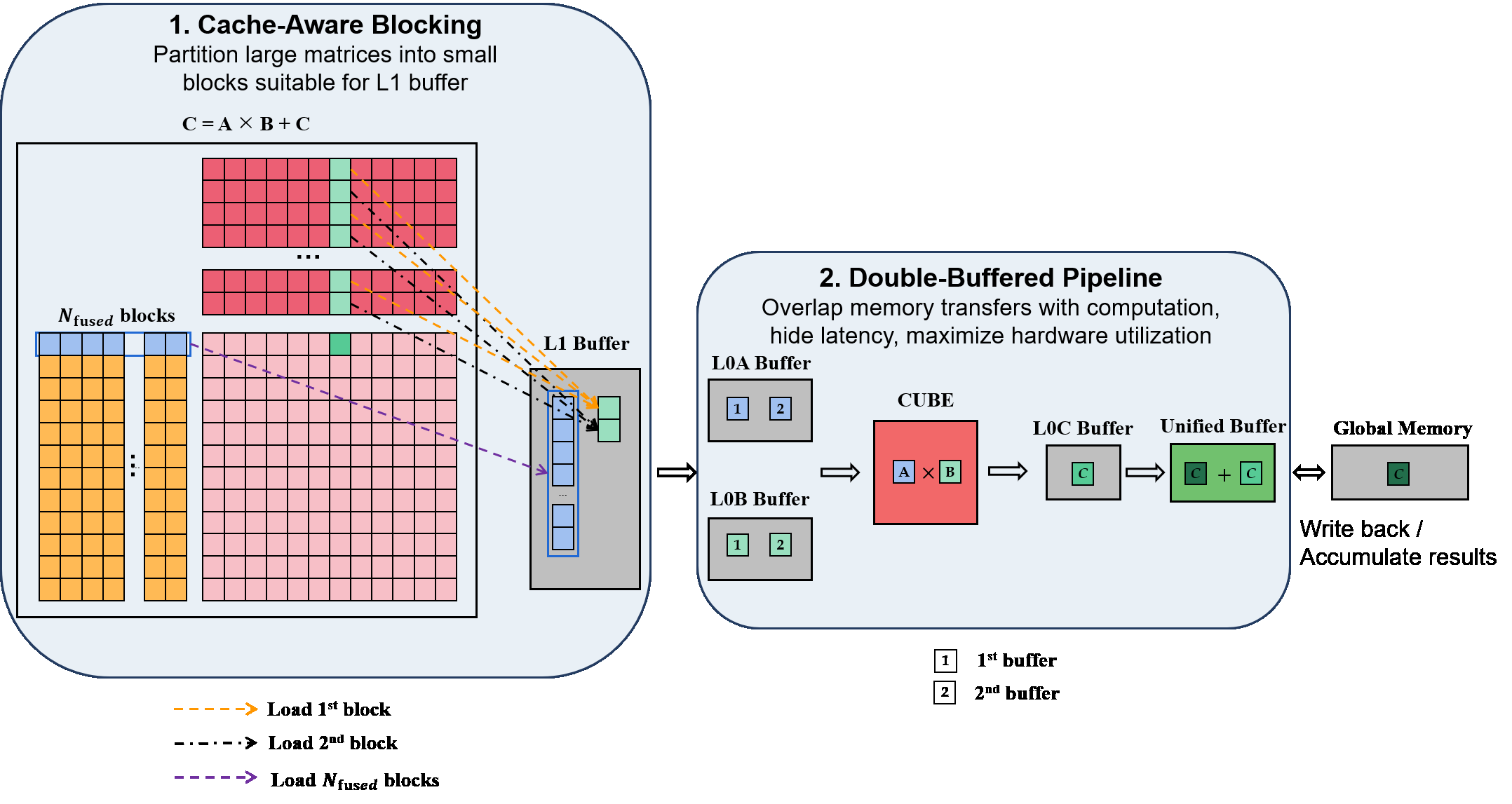}
\caption{Matrix blocking based on L1 cache reuse}
\Description{Illustration of the cache-aware matrix blocking strategy. Blocks of matrix A remain resident in L1 cache for reuse, while blocks of matrix B are streamed through double-buffered L1 regions to overlap data movement and computation.}
\label{matrix_blocking}
\end{figure}

These principles aim to minimize data movement between main memory and L1 while maintaining a steady supply of operands to the cube units. Fig.~\ref{matrix_blocking} shows the block layout: $A$ blocks are stored in row-major order in L1, and $B$ blocks are streamed column-wise into alternating L1 buffers.

\paragraph{L1 Reuse Quantification and Traffic Proxy}

The degree of reuse is quantified by the number of $A$ blocks that can be held in L1 simultaneously:
\begin{equation}
N_{\text{fused}} = \text{int}\!\left(\frac{L1 - 2 b_k b_n}{b_m b_k}\right)
= \text{int}\!\left(\frac{L1}{b_m b_k} - 2 \frac{b_n}{b_m}\right)
= f \frac{L1}{b_m b_k},
\label{N_fused}
\end{equation}
where $b_m$ and $b_k$ denote the block height and width of $A$, and $f$ ($0.92 \leq f \leq 1$ in our experiments) accounts for the correction from $\frac{b_n}{b_m} \sim O(1)$ and the floor operation. A larger $N_{\text{fused}}$ allows more resident $A$ tiles to share the streamed operand under the implemented loop ordering.

For the original parameter sweep, we use the following normalized traffic proxy:
\begin{equation}
\left\{
\begin{aligned}
A_\mathrm{r} & = m k, \\
B_\mathrm{r} & = \frac{mkn}{N_{\mathrm{core}}\, b_m}, \\
C_\mathrm{rw} & = \frac{2 m k n\, b_m}{f\, L1}, \\
\mathrm{data}_\mathrm{proxy} & = A_\mathrm{r} + B_\mathrm{r} + C_\mathrm{rw},
\end{aligned}
\right.
\label{combined_formula}
\end{equation}
where $m,k,n$ denote the matrix dimensions, $b_m,b_k,b_n$ the block sizes, $N_{\mathrm{core}}=32$ the number of AI cores, and $f$ the L1 fusion efficiency factor. The three terms use the sharing assumptions of the implemented schedule and are useful for ranking candidate blocks, but they are not a complete physical aggregate-traffic identity: the exact byte count also depends on inter-core replication, loop ordering, and the placement of partial results. We therefore treat Eq.~\ref{combined_formula} as a heuristic proxy rather than as a proof of a bottleneck.

\paragraph{Normalized Intensity and First-Level Roofline Proxy}
From Eq.~\eqref{N_fused} and Eq.~\eqref{combined_formula}, we define the normalized intensity proxy
\begin{equation}
\mathrm{OI}_{\mathrm{proxy}}
= \frac{2 m n k}
       {\,s_A A_\mathrm{r} + s_B B_\mathrm{r} + s_C C_\mathrm{rw}\,},
\label{eq:oi_roof}
\end{equation}
where $2mnk$ is the FP32-equivalent FLOP count of a single GEMM and $s_A,s_B,s_C$ are the element sizes used by the proxy (in our setting $s_A\!=\!s_B\!=\!s_C\!=\!4$). A corresponding first-level ceiling is plotted as
\begin{equation}
P_{\text{proxy}} \;=\; \min\!\big(\;P_{\text{peak}}\;,\;\beta_{\text{mem}\rightarrow \text{L1}} \cdot \mathrm{OI}_{\text{proxy}}\;\big),
\label{eq:roofline_bound}
\end{equation}
where $P_{\text{peak}}$ is the FP32-equivalent compute ceiling and $\beta_{\text{mem}\rightarrow \text{L1}}$ is the sustained main-memory$\rightarrow$L1 bandwidth. Because Eq.~\ref{combined_formula} is a scheduling proxy rather than a complete aggregate byte count, $P_{\text{proxy}}$ is used only as a qualitative first-level diagnostic. The measured blocking and pipeline ablations remain the basis for the performance conclusions.

\paragraph{Hardware Block-Size Constraints}
Block sizes are subject to hardware constraints arising from cube computation alignment and buffer capacities:
\begin{equation}
\begin{cases}
    b_m, b_k, b_n \equiv 0 \pmod{16} & \text{(cube alignment)} \\
    b_m \times b_k \leq 64 \times 256 & \text{(L0A capacity)} \\
    b_k \times b_n \leq 64 \times 256 & \text{(L0B capacity)} \\
    b_m \times b_n \times 6 \leq 248 \times 1024 & \text{(L0C and UB capacity)}
\end{cases}
\label{block_limit}
\end{equation}

\paragraph{Impact on L1 Fusion and the Initial $b_m$ Candidate}
\begin{figure}[!t]
    \centering
    \begin{subfigure}{0.4\textwidth}
        \centering
        \includegraphics[width=0.9\linewidth]{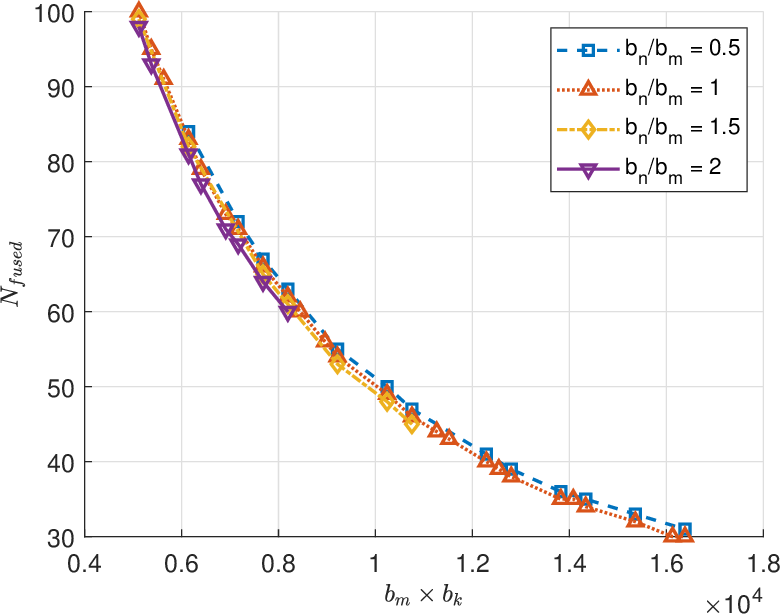}
        \caption{$N_\text{fused}$ vs $b_m \times b_k$}
    \end{subfigure}
    \begin{subfigure}{0.4\textwidth}
        \centering
        \includegraphics[width=0.9\linewidth]{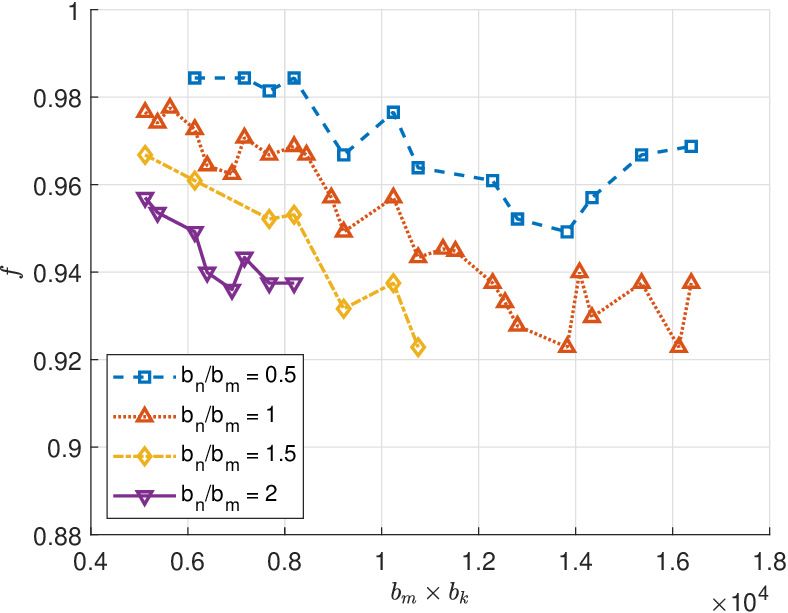}
        \caption{$f$ vs $b_m \times b_k$}
    \end{subfigure}
    \caption{Impact of blocking size on $N_\text{fused}$ and $f$}
    \Description{Two plots showing how matrix block dimensions affect cache fusion efficiency. The first plot shows the number of fused matrix blocks as block size changes. The second plot shows the corresponding fusion efficiency factor across different blocking configurations.}
    \label{impact_nfused_f}
\end{figure}

Exploring the feasible $(b_m,b_k,b_n)$ space under Eq.~\ref{block_limit} shows that $N_{\text{fused}}$ decreases with $b_m b_k$, while $f$ remains high for $0.5 \le b_n/b_m \le 2$ (Fig.~\ref{impact_nfused_f}). The stationary point of the proxy in Eq.~\ref{combined_formula} is
\[
\frac{\partial}{\partial b_m}\!\left(\frac{mkn}{N_{\mathrm{core}}\,b_m} \;+\; \frac{2mkn\,b_m}{f\,L1}\right)=0
\;\Rightarrow\;
b_{m,\mathrm{proxy}} \;=\; \sqrt{\frac{f\,L1}{2\,N_{\mathrm{core}}}}.
\]
On Ascend 910A this expression gives an initial candidate near the feasible multiple $b_m=96$. Because the proxy does not account for all replication, L0 utilization, DMA setup, and scheduling effects, it is not asserted to be a physical optimum. We therefore search the feasible block space experimentally; Sec.~\ref{performance} reports both the $b_m=96$ baseline and the measured best configuration.

\subsection{Double-Buffered Pipeline}
\label{double_buffering}

\begin{figure}[!t]
\centering
\includegraphics[width=4.5in]{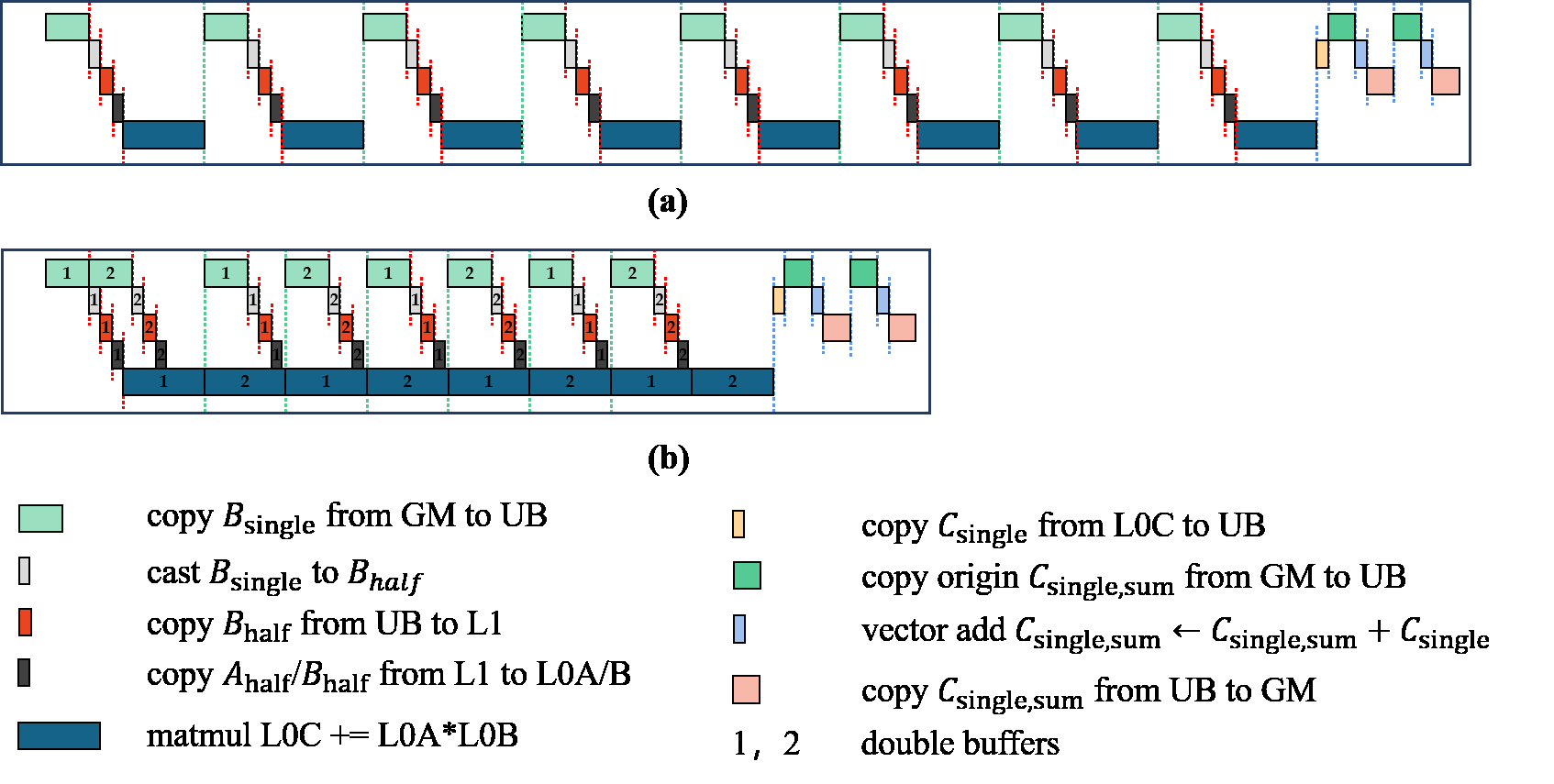}
\caption{Comparison of single- and double-buffered pipelines based on L1 reuse}
\Description{Timeline comparison between single-buffered and double-buffered GEMM execution pipelines. The double-buffered design overlaps memory transfers and FP16 cube computation to reduce idle time and improve utilization.}
\label{pipeline}
\end{figure}

While cache-aware blocking reduces memory traffic, the pipeline structure still constrains throughput: in a single-buffered design, each GEMM step must wait for data to be fully loaded into L1 before computation, leaving the cube units idle (Fig.~\ref{pipeline} (a)). To overcome this limitation, we implement a double-buffered pipeline that overlaps data movement with computation across L1, L0A, and L0B, thereby maximizing concurrency (Fig.~\ref{pipeline} (b)).

The key idea is to maintain two L1 buffers for matrix $B$ blocks: while one buffer feeds the cube unit, the other prefetches the next block from global memory. This ensures that when the current block computation finishes, the next block is already staged in L1. Matrix $A$ blocks, already optimized for reuse, remain resident in L1 across multiple inner-loop iterations, following the cache-aware blocking strategy in Sec.~\ref{sec:cache_blocking}.

Let $T_{\mathrm{comp}}$ and $T_{\mathrm{mem}}$ denote the computation and transfer time per block. A single-buffered pipeline requires $T_{\mathrm{comp}} + T_{\mathrm{mem}}$ per iteration, whereas the double-buffered design approaches $\max(T_{\mathrm{comp}}, T_{\mathrm{mem}})$, effectively hiding memory latency whenever $T_{\mathrm{comp}} \gtrsim T_{\mathrm{mem}}$. In practice, overheads such as DMA setup and synchronization may prevent the full hiding of memory latency, forming a more practical performance bound of $T_{\mathrm{comp}} + \alpha T_{\mathrm{mem}}$, where $\alpha$ represents the non-overlapped fraction of memory time.

The decomposition in Eq.~\ref{recover_eqn} produces three structurally similar GEMM terms, each of which can be computed independently. 
To fully exploit the parallelism of the Ascend architecture and hide memory latency, we implement each term using a double-buffered GEMM pipeline that overlaps data transfers with computation. This approach ensures that while one set of matrix blocks is being processed by the FP16 cube units, the next set is prefetched into L1, thereby keeping the compute units busy and minimizing idle cycles. Algorithm~\ref{alg:double_buffer} outlines the generic procedure for one such term, which is directly applicable to $A_{\text{half}}B_{\text{half}}$, $R_{A,\text{half}}B_{\text{half}}/s_f$, and $A_{\text{half}}R_{B,\text{half}}/s_f$.

\begin{algorithm}[!t]
\floatname{algorithm}{Algorithm}
\caption{Double-buffered GEMM for one decomposition term in SGEMM-cube}
\label{alg:double_buffer}
\footnotesize
\KwIn{Matrix term $(A, B)$ partitioned into $M\times K$ and $K\times N$ blocks; scaling factor $sf$}
\KwOut{Accumulated result $C \leftarrow A \times B$}
\ForEach{block row of $A$ in parallel over $N_{\text{core}}$ cores}{
    \ForEach{group of $N_{\text{fused}}$ $A$-blocks}{
        \tcp{Stage $A$ blocks into L1 (once per group)}
        \textbf{prefetch} FP16 component blocks of A into L1\;
        \ForEach{column block of $B$}{
            \tcp{Double-buffered prefetch of $B$ blocks}
            \While{computing on buffer $p$}{
                \textbf{prefetch} next $B$ block to buffer $q$\;
            }
            \tcp{GEMM computation}
            $C_{\mathrm{blk}} \leftarrow A_{\mathrm{blk}} \times B_{\mathrm{blk}} + C_{\mathrm{blk}}$\;
            \tcp{Write-back if UB capacity reached}
            \textbf{accumulate and store} $C_{\mathrm{blk}}$ to global memory\;
        }
    }
}
\end{algorithm}

\section{Experimental Results}

This section evaluates SGEMM-cube along two axes: (i) numerical accuracy of the FP32 approximation (Sec.~\ref{error}), and (ii) sustained throughput and scalability on Ascend NPUs (Sec.~\ref{performance}). We compare against FP16 HGEMM and FP32 baselines and report how the proposed system-level optimizations (Sec.~\ref{system_optimizations}) translate into end-to-end performance gains.

\subsection{Experimental Setup}

For each accuracy experiment, matrix entries are independently sampled from either a symmetric uniform distribution
$U[-2^e,2^e]$ or a non-negative uniform distribution $U[0,2^e]$, where $e$ denotes the FP32 offset exponent. Unless otherwise stated, each configuration is repeated five times with different random seeds, and we report the average relative error. The tested scaling exponents are $s_b\in\{0,6,12\}$, corresponding to $s_f\in\{1,2^6,2^{12}\}$.

For accuracy evaluation, FP64 DGEMM is used as the reference result. FP32 OpenBLAS SGEMM, FP16 HGEMM, and SGEMM-cube are compared against this FP64 reference. For performance evaluation, throughput is computed as $2mnk/t$, where $t$ is the measured kernel execution time. For SGEMM-cube, we report FP32-equivalent throughput following the convention that one approximate FP32 GEMM requires three dominant FP16 GEMM operations.

\textbf{Platforms}
Three machines are involved. Note that all SGEMM-cube kernels run on the Ascend NPU; the Kunpeng CPU listed below is only the host and the CPU reference.
\begin{itemize}
    \item \textbf{Target: Ascend 910A NPU.} 32 AI cores at 1\,GHz, 256\,TFLOP/s FP16 Cube throughput, 1.2\,TB/s main-memory bandwidth, and \emph{no} native FP32 matrix units. This is the platform on which SGEMM-cube and the FP16 HGEMM baseline are measured. Architectural details are given in Fig.~\ref{AI_core_architecture} and Sec.~\ref{sec:architecture}.
    \item \textbf{FP32 NPU reference: Ascend 910B3.} 20 AI cores at 1.8\,GHz, half the L1 capacity per core relative to the 910A, doubled main memory, and 1.6\,TB/s bandwidth. Unlike the 910A it supports native FP32 GEMM, with a theoretical peak of 73.73\,TFLOP/s. We run the vendor CANN SGEMM here as an FP32 hardware reference.
    \item \textbf{Host and CPU reference: Kunpeng 920.} 192 64-bit ARM cores across 4 sockets (48 cores/socket) in 8 NUMA nodes, each core with 64\,KB L1, 512\,KB L2 and 48\,MB shared L3. This CPU hosts the 910A node and is used only to run the FP64 DGEMM reference and the OpenBLAS FP32 SGEMM baseline; no part of SGEMM-cube executes on it.
\end{itemize}

\textbf{Baselines and metrics}
For accuracy, we use FP64 DGEMM as the ground-truth reference and FP32 SGEMM (OpenBLAS) as a software baseline. On NPUs, we compare FP16 HGEMM and SGEMM-cube (elementwise/termwise variants). The relative error is defined in Eq.~\ref{err}. For performance, we report kernel throughput (TFLOPs) as $\frac{2mkn}{\text{time}}$ on square or rectangular matrices, 
where $2mkn$ is the total number of floating-point operations and $\text{time}$ is the kernel execution time. Unless noted, the termwise accumulation and cache-aware/double-buffered execution are enabled when evaluating SGEMM-cube.

\textbf{Input generation}
To probe underflow-prone regimes and cancellation effects, we follow two sampling strategies for FP32 inputs: (a) symmetric ranges $[-2^{\text{offset exponent}}, 2^{\text{offset exponent}}]$ and (b) non-negative ranges $[0, 2^{\text{offset exponent}}]$ (Fig.~\ref{scaling_bit_error}). We evaluate scaling exponents $s_b\in\{0,6,12\}$ (refer to Sec.~\ref{scaling}).

\subsection{Accuracy Evaluation}
\label{error}

We report the Frobenius-norm relative error
\begin{equation}
\text{err}
=
\frac{\|C_\text{true}-C_\text{calculated}\|_F}
     {\|C_\text{true}\|_F}.
\label{err}
\end{equation}
where $C_{\text{true}}$ is obtained via FP64 DGEMM on the Kunpeng CPU, and $C_{\text{calculated}}$ covers SGEMM-cube (both variants), FP16 HGEMM, and FP32 OpenBLAS SGEMM.

A Frobenius-norm measure aggregates errors across entries and can hide a small number of badly computed entries. We do not measure the componentwise error directly. Under the range and rounding assumptions stated in Sec.~\ref{sec:err_bound}, Eq.~\ref{eq:main_bound} bounds the termwise result entry by entry; the measurements below report the realized normwise error rather than the worst-case coefficient. The two views are complementary. What is still missing is an empirical max-norm study under adversarial rather than random inputs, which we leave to future work (Sec.~\ref{sec:limitations}).

\begin{figure}[!t]
    \centering
    \begin{subfigure}{0.49\textwidth}
        \centering
        \includegraphics[width=0.9\linewidth]{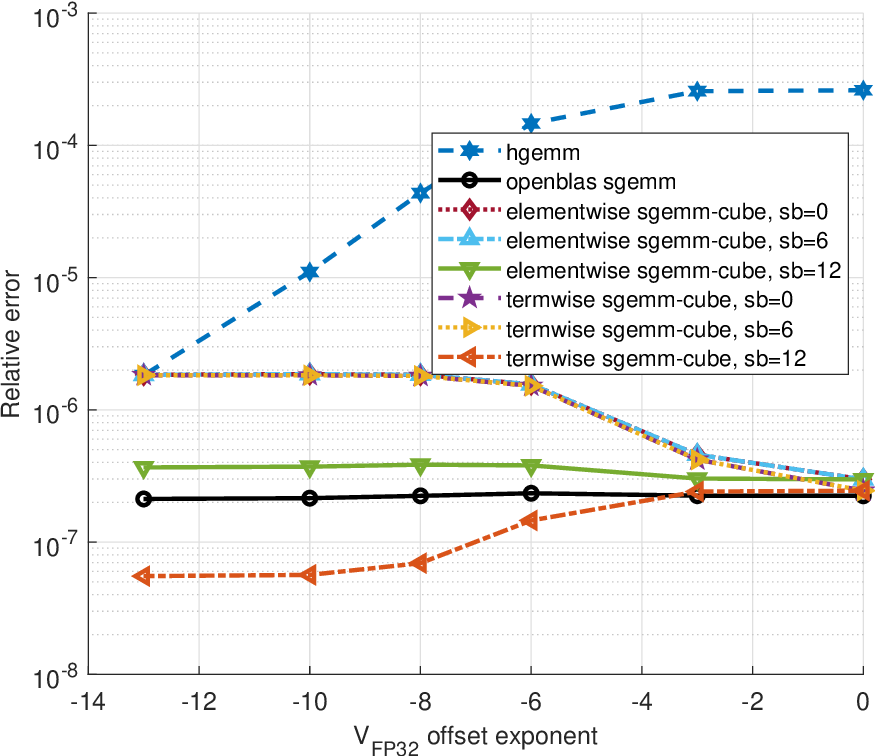}
        \caption{Uniform sampling from $[-2^{\text{offset exponent}}, 2^{\text{offset exponent}}]$}
    \end{subfigure}
    \begin{subfigure}{0.49\textwidth}
        \centering
        \includegraphics[width=0.9\linewidth]{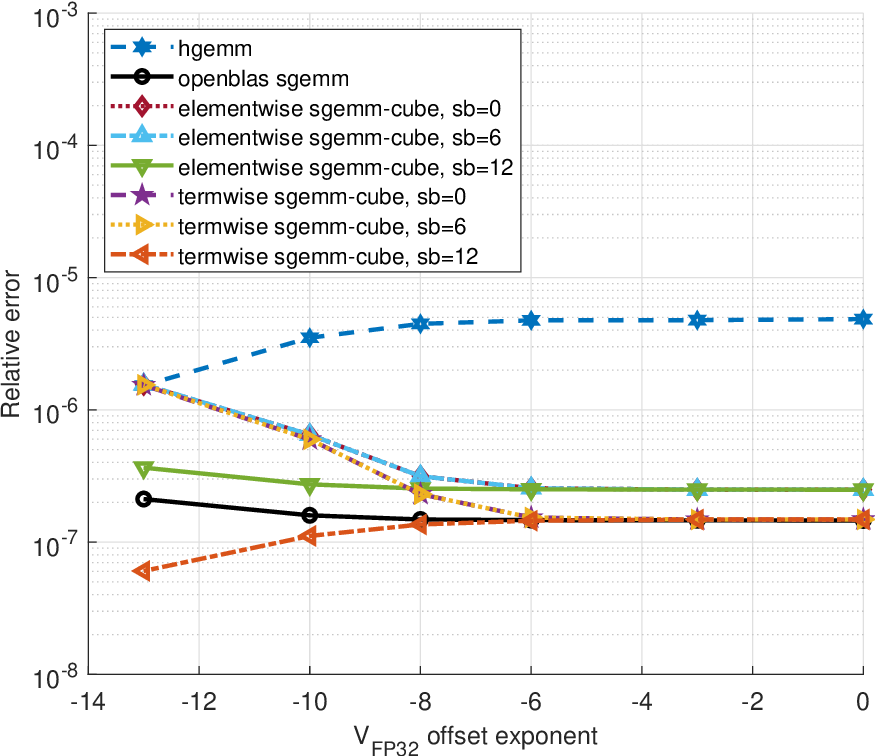}
        \caption{Uniform sampling from $[0, 2^{\text{offset exponent}}]$}
    \end{subfigure}
    \caption{Relative error vs. offset exponent under different input ranges}
    \Description{Plots of relative error versus FP32 offset exponent under two input sampling strategies. The left plot uses symmetric sampling around zero and highlights cancellation effects. The right plot uses non-negative sampling and shows improved numerical stability under residual scaling.}
    \label{scaling_bit_error}
\end{figure}

Fig.~\ref{scaling_bit_error} sweeps the FP32 offset exponent under the two sampling regimes. With symmetric sampling (Fig.~\ref{scaling_bit_error} (a)), destructive cancellation can depress $\|C_{\text{true}}\|_F$, amplifying the reported relative error (Eq.~\ref{err}); this effect is absent for non-negative sampling (Fig.~\ref{scaling_bit_error} (b)). Across both settings, FP16 HGEMM exhibits the highest error ($\sim 10^{-4}$). Without scaling ($s_b=0$), elementwise and termwise SGEMM-cube behave similarly, with termwise slightly better near zero offset exponent, but both trail FP32 SGEMM, highlighting the necessity of residual amplification. Setting $s_b=12$ improves accuracy by $1\sim2$ orders of magnitude in the evaluated low-exponent regime, while $s_b=6$ is insufficient; as Sec.~\ref{scaling} states, this choice is valid only within the corresponding upper-range condition.

Once scaling is applied the two accumulation strategies separate. With $s_b=12$, elementwise SGEMM-cube is comparable to the baseline SGEMM while termwise SGEMM-cube surpasses it in the small-exponent region. Equation~\ref{eq:main_bound} models the termwise organization only; the elementwise organization incurs additional block-level additions whose effect is measured here rather than covered by that bound. Accumulating the correction streams $\sum_g R_A^{[g]}B^{[g]}$ and $\sum_g A^{[g]}R_B^{[g]}$ separately (Eq.~\ref{eq:termwise}) sums the small contributions among themselves at FP32 precision and combines them with the dominant product $\sum_g A^{[g]}B^{[g]}$ only once. This grouping reduces repeated swamping of the correction terms and explains the empirical advantage over the tested OpenBLAS FP32 baseline. It is not a general claim of superiority over every FP32 GEMM accumulation order.

\begin{figure}[!t]
    \centering
    \begin{subfigure}{0.32\textwidth}
        \centering
        \includegraphics[width=0.9\linewidth]{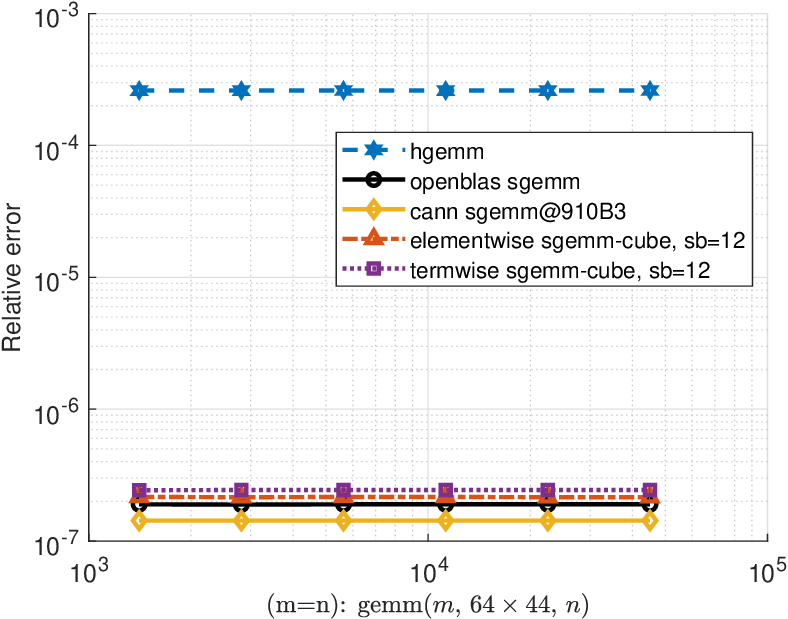}
        \caption{Relative error vs. $m$ and $n$}
    \end{subfigure}
    \begin{subfigure}{0.32\textwidth}
        \centering
        \includegraphics[width=0.9\linewidth]{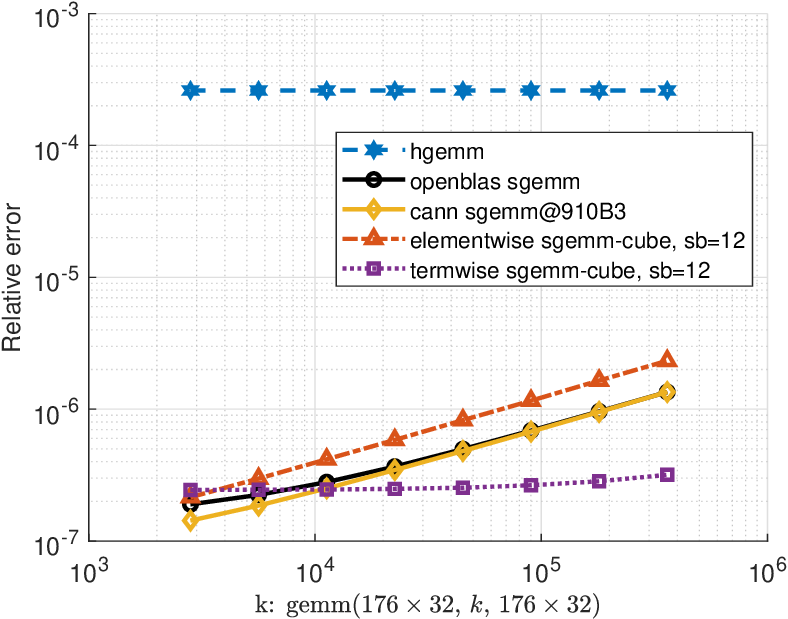}
        \caption{Relative error vs. $k$}
    \end{subfigure}
    \begin{subfigure}{0.32\textwidth}
        \centering
        \includegraphics[width=0.9\linewidth]{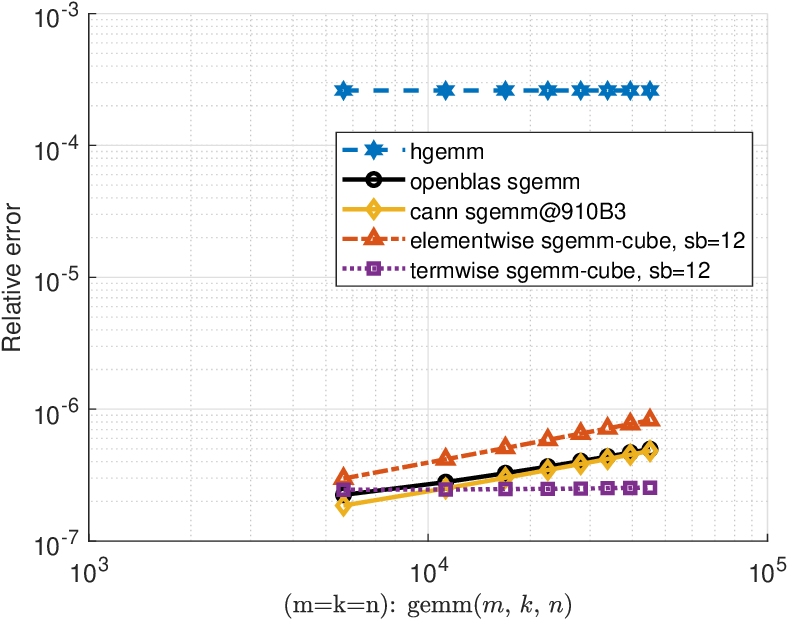}
        \caption{Relative error vs. $m$, $k$, and $n$}
    \end{subfigure}    
    \caption{Relative error vs. input matrix sizes}
    \Description{Plots showing how relative error changes with matrix dimensions. The experiments separately vary $m$ and $n$, vary $k$, and jointly vary all three dimensions to evaluate accumulation stability and scaling behavior.}
    \label{mkn_error}
\end{figure}

Fig.~\ref{mkn_error} examines how error scales with matrix size at offset exponent $=0$. Varying $m$ and $n$ with fixed $k=64\times 44$ (Fig.~\ref{mkn_error} (a)) leaves the error nearly unchanged since accumulation depth is governed by $k$. Increasing $k$ (Fig.~\ref{mkn_error} (b) and (c)) stresses summation stability; here, \emph{termwise} SGEMM-cube consistently outperforms FP32 OpenBLAS SGEMM and the elementwise variant, aligning with our design that sums smaller-magnitude contributions first.

\subsection{Performance Evaluation}
\label{performance}

\begin{figure}[!t]
\centering
\includegraphics[width=3.5in]{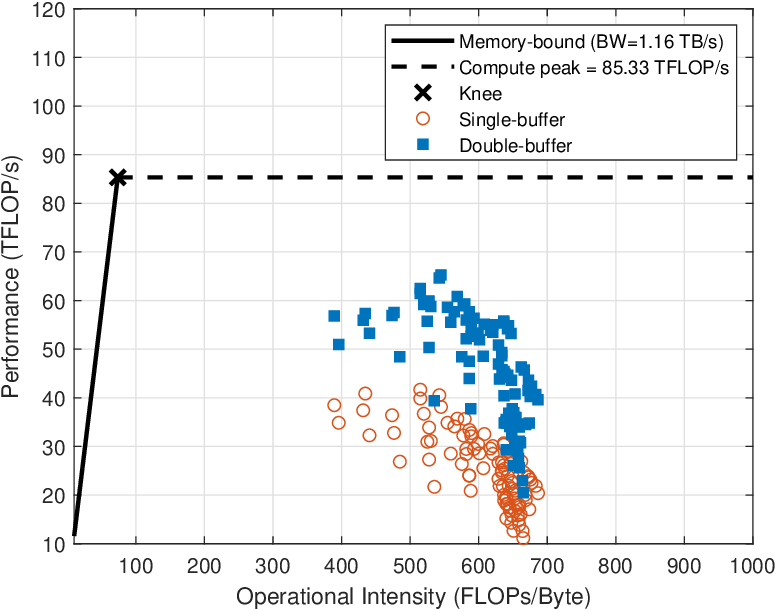}
\caption{First-level intensity proxy for the single-buffer and double-buffer schemes on Ascend~910A. The plot is retained as a qualitative view of the parameter sweep; the measured throughput improvement from double buffering does not rely on a compute-bound classification.}
\label{roofline_combined_linear}
\Description{A first-level intensity proxy comparing single-buffered and double-buffered SGEMM-cube implementations on Ascend 910A. The measured points show the throughput improvement from double buffering, while the proxy is not used as a complete hierarchical bottleneck model.}
\end{figure}

\paragraph{Roofline-Guided Observations}
Using Eq.~\ref{N_fused}, Eq.~\ref{combined_formula}, and Eq.~\ref{eq:oi_roof}, we compute the normalized intensity proxy for each $(b_m,b_k,b_n)$ configuration and plot the first-level ceiling of Eq.~\ref{eq:roofline_bound}. The FP32-equivalent compute ceiling is defined as one third of the hardware's native FP16 peak, accounting for the three primary matrix multiplications in Eq.~\ref{recover_eqn}. Figure~\ref{roofline_combined_linear} overlays the measured single-buffer and double-buffer performance on this proxy. Because Eq.~\ref{combined_formula} is not a complete aggregate byte model, the figure is used only as a qualitative view of the parameter sweep and does not by itself prove that every point is compute-bound. The robust observation is empirical: double buffering improves realized throughput by overlapping data transfer and computation, while both implementations remain below the three-GEMM compute ceiling.

Figure~\ref{roofline_combined_linear} has two additional limitations. It represents only the main-memory\,$\to$\,L1 path, whereas a complete hierarchical analysis would add separate L1\,$\to$\,L0A/L0B and L0C\,$\to$\,UB ceilings and would count aggregate traffic consistently across cores. Its linear axes also differ from the conventional log--log Roofline presentation. We therefore do not use this figure to localize the residual gap to the compute ceiling; instruction-level and multi-level traffic measurements would be required for that purpose.

\begin{figure}[!t]
    \centering
    \begin{subfigure}{0.49\textwidth}
        \centering
        \includegraphics[width=0.9\linewidth]{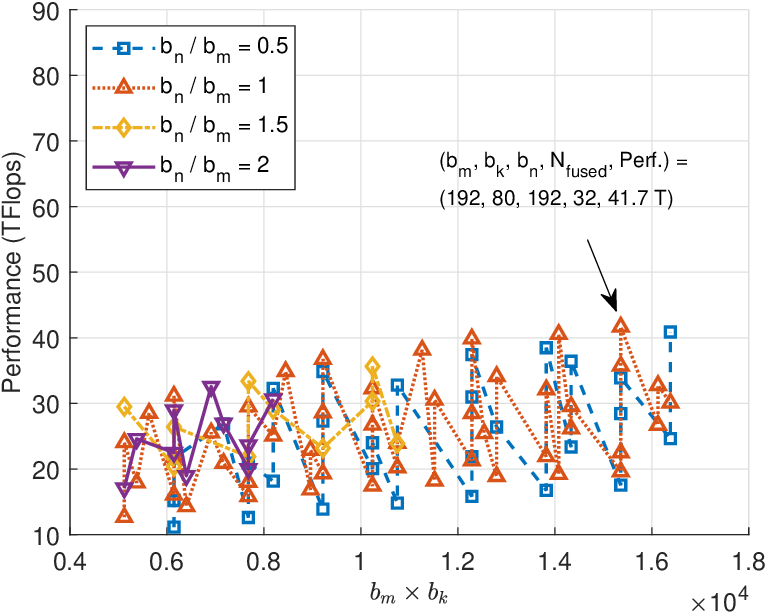}
        \caption{Performance of single-buffered pipeline}
    \end{subfigure}
    \begin{subfigure}{0.49\textwidth}
        \centering
        \includegraphics[width=0.9\linewidth]{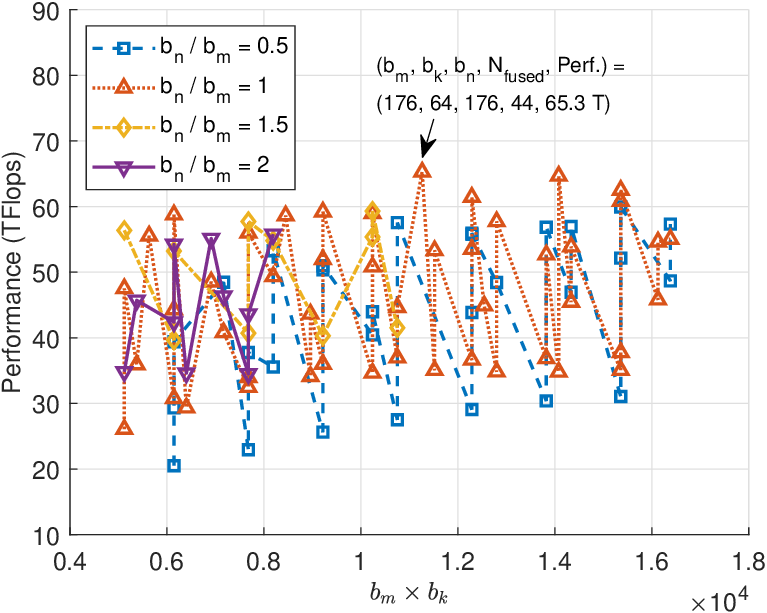}
        \caption{Performance of double-buffered pipeline}
    \end{subfigure}
    \caption{Performance impact of matrix blocking with L1 reuse}
    \Description{Performance heatmaps comparing single-buffered and double-buffered SGEMM-cube pipelines under different matrix block sizes. Double buffering consistently achieves higher throughput by overlapping data movement with computation.}
    \label{perf_pipeline}
\end{figure}

\paragraph{Pipeline Ablation (Single vs. Double buffering)}
Fig.~\ref{perf_pipeline} sweeps block sizes and contrasts pipeline designs on 910A. The single-buffered pipeline peaks at 41.7\,TFLOP/s (Fig.~\ref{perf_pipeline}(a)). Enabling double buffering raises the peak to 65.3\,TFLOP/s (Fig.~\ref{perf_pipeline}(b)), a 57\% gain by overlapping DMA transfers with compute and reducing stalls. Normalized to the FP32-equivalent compute roof on 910A (raw FP16 peak $256$\,TFLOP/s), this attains 77\% of $256/3{=}85.3$\,TFLOP/s. Low points at small blocks are explained by poor L0A/L0B utilization (pipeline bubbles) not explicitly captured by Eq.~\ref{block_limit}. The best configuration $(b_m,b_k,b_n,N_{\text{fused}}){=}(176,64,176,44)$ attains the maximum throughput.

\begin{figure}[!t]
    \centering
    \begin{subfigure}{0.32\textwidth}
        \centering
        \includegraphics[width=0.9\linewidth]{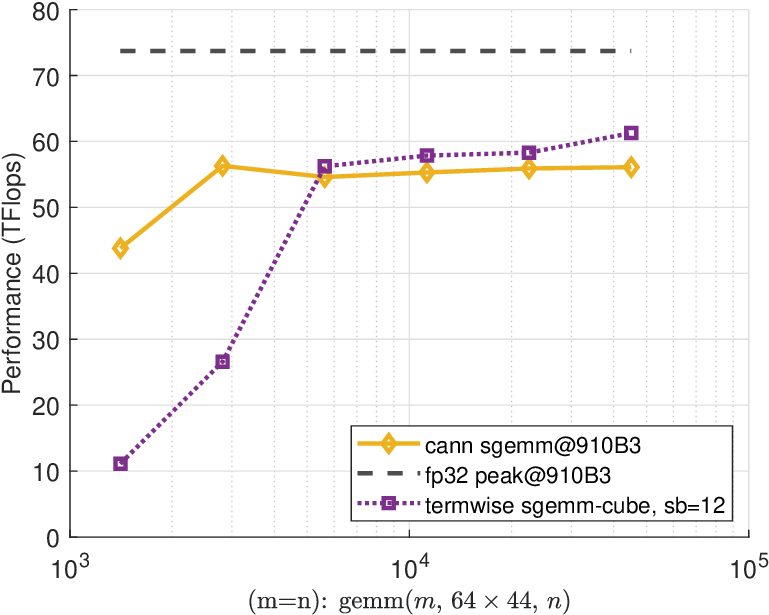}
        \caption{Performance vs. $m$ and $n$}
    \end{subfigure}
    \begin{subfigure}{0.32\textwidth}
        \centering
        \includegraphics[width=0.9\linewidth]{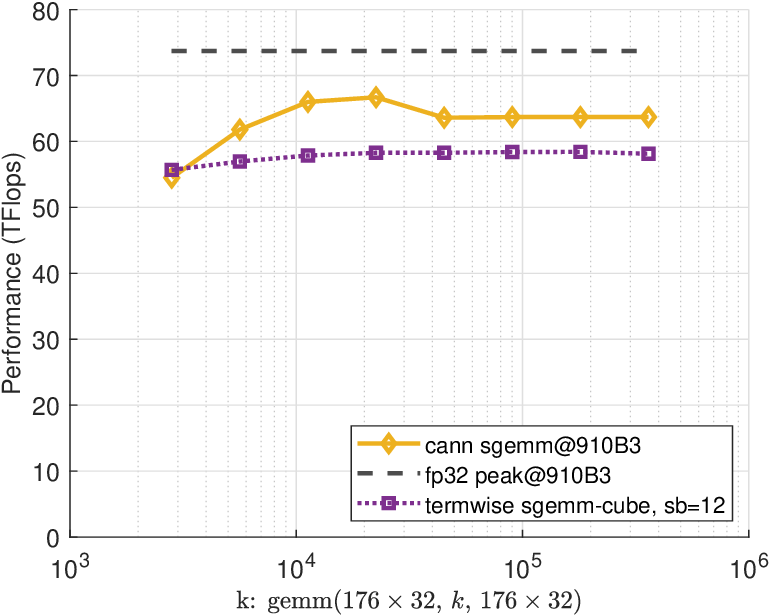}
        \caption{Performance vs. $k$}
    \end{subfigure}
    \begin{subfigure}{0.32\textwidth}
        \centering
        \includegraphics[width=0.9\linewidth]{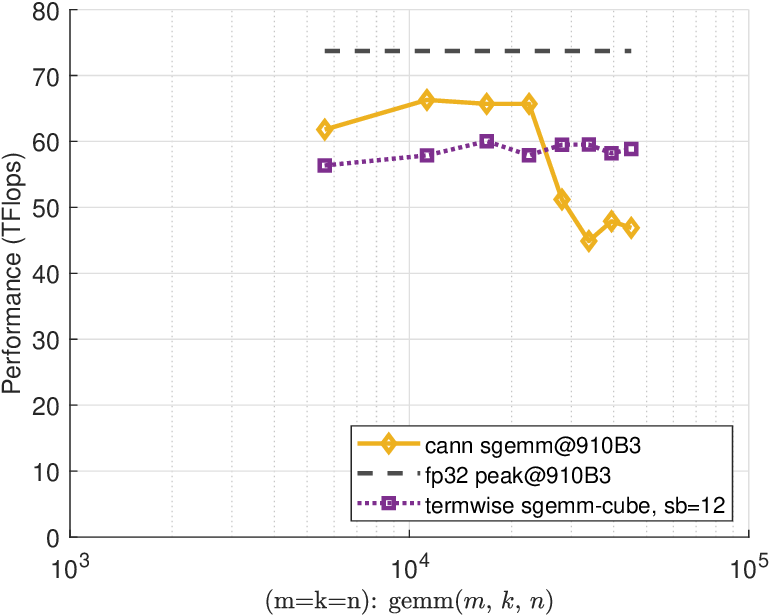}
        \caption{Performance vs. $m$, $k$, and $n$}
    \end{subfigure}    
    \caption{Performance vs. input matrix sizes}
    \Description{Performance scaling results for SGEMM-cube across different matrix dimensions. The plots evaluate throughput sensitivity to m and n, k, and jointly scaled matrix sizes on Ascend NPUs.}
    \label{mkn_perf}
\end{figure}

\paragraph{Size scaling and cross-platform comparison}
Fig.~\ref{mkn_perf} studies scaling. Increasing $m,n$ pushes throughput past 60\,TFLOP/s on 910A (Fig.~\ref{mkn_perf}(a)), slightly surpassing CANN FP32 SGEMM on 910B3 despite 910B3 having native FP32 units. As $k$ grows (Fig.~\ref{mkn_perf}(b)), both SGEMM-cube@910A and CANN FP32@910B3 remain stable; SGEMM-cube reaches $\sim$60\,TFLOP/s versus $\sim$63\,TFLOP/s on 910B3. When jointly scaling $m,k,n$ (Fig.~\ref{mkn_perf}(c)), CANN FP32@910B3 degrades at very large sizes, whereas SGEMM-cube@910A maintains stable performance and eventually surpasses it, indicating that L1-aware blocking and double buffering sustain utilization even when matrices outgrow typical cache capacities.

\subsection{Discussion: What Transfers to Other Hardware}

Ascend NPUs are not yet widely available outside a small number of installations. We should be clear about which of our results depend on that hardware and which do not.

\emph{Hardware-independent under the stated arithmetic model.} The decomposition and range inequalities of Sec.~\ref{sec:accuracy_analysis} are format-level results, but the concrete error bound of Sec.~\ref{sec:err_bound} requires FP16 products to be exactly representable in FP32 and the three termwise accumulations and final additions to follow the stated RN model. The scale is range-dependent: $s_b=12$ is valid only under the upper-exponent condition in Eq.~\ref{sf_bounds}, while $s_b=11$ is the conservative fixed choice over the full normal-FP16 high-component range. The arithmetic motivation for termwise accumulation transfers to other devices, but its measured advantage must be re-evaluated for their reduction order. Under the assumed model, the exact low-low coefficient is dominated by $\gamma_k$ for $k\geq5$, and the complete bound is within a factor of two of the classical FP32 bound for $k\geq15$.

\emph{Hardware-dependent.} The L1-capacity relation in Eq.~\ref{N_fused}, the constraints of Eq.~\ref{block_limit}, the traffic proxy in Eq.~\ref{combined_formula}, and every measured throughput are specific to the 910A. The general strategy transfers---keep one operand resident and double-buffer the streamed operand---but aggregate traffic, feasible tile sizes, and the best $b_m$ must be re-derived and re-measured for each target. On GPUs and TPUs the relevant capacities are shared memory and registers rather than L1/L0A/L0B/UB, and hardware caching changes the cost of a poor choice.

One systems observation is worth carrying across. Double buffering alone accounts for a 57\% throughput gain here (Sec.~\ref{performance}), lifting the kernel from 41.7 to 65.3\,TFLOP/s without changing its operational intensity. Where no hardware prefetcher or cache is available to hide transfer latency, explicit pipelining is not a late-stage tuning step but a first-order design decision, and an emulation kernel ported to such a device ought to be structured around it from the start. We have no matched measurements on a cache-based accelerator and so offer no quantitative comparison.

\section{Limitations}
\label{sec:limitations}

\emph{Numerical scope.} SGEMM-cube is not a bit-exact FP32 emulation. The first-order form of the termwise bound contains the scheme-specific $14u_{32}$ contribution in Eq.~\ref{eq:main_bound_first_order}, while Eq.~\ref{eq:main_bound} retains the higher-order terms. Nor does the method cover the full FP32 dynamic range: inputs outside the FP16 representable window need additional exponent management or input scaling, which the current version does not implement. The experiments use the fixed scale $s_f=2^{12}$ within the stated exponent window; covering the full normal-FP16 high-component range requires $s_b=11$, dynamic scaling, or an overflow guard. Our analysis also assumes that the Cube unit accumulates FP16 products into FP32 with round-to-nearest. Characterizing the unit's internal rounding independently, along the lines of the Tensor Core study of Fasi et al.~\cite{fasi2021numerical}, would put that assumption on firmer ground.

\emph{Evaluation scope.} The accuracy study reports Frobenius-norm relative errors on randomly sampled inputs. Equation~\ref{eq:main_bound} bounds the termwise componentwise error a priori under the stated arithmetic assumptions, but we never measure it, and a normwise average can hide a small number of badly computed entries. A max-norm study using adversarially constructed, cancellation-heavy inputs is the most useful experiment missing from this paper. The first-level proxy in Fig.~\ref{roofline_combined_linear} is also incomplete: it uses the schedule-normalized traffic expression of Eq.~\ref{combined_formula}, covers only the main-memory\,$\to$\,L1 path, and uses linear axes. A consistent aggregate, hierarchical Roofline analysis would be needed to localize the residual gap to the compute ceiling.

\emph{Performance scope.} Main-memory bandwidth alone does not explain the gap to the three-GEMM FP32-equivalent peak. DMA setup overhead, synchronization cost, instruction-scheduling overhead, imperfect L0A/L0B utilization for small tiles, and the conversion and reconstruction work around the Cube GEMMs are all plausible contributors. Separating them requires instruction- and pipeline-level profiling that we have not done.

\emph{Portability and reproducibility.} The implementation targets the Ascend 910A, and that hardware is not broadly available for rent or through research clusters, so third parties cannot easily reproduce the performance results. We have tried to keep the reproducible parts of this work independent of the hardware. The analysis of Sec.~\ref{sec:accuracy_analysis} and Sec.~\ref{sec:err_bound} can be checked on paper, and the accuracy experiments of Sec.~\ref{error} can be replicated on any platform by emulating FP16 rounding in software. The blocking constants and throughput figures cannot.

\section{Conclusion}
This paper presented SGEMM-cube, a precision-recovery approximation for FP32 GEMM on Ascend NPUs with FP16 Cube units. It follows the Ootomo-style two-word splitting strategy, a fixed-length multiword scheme distinct from the error-free Ozaki scheme, and rebuilds the matrix product from three dominant FP16 GEMM terms with FP32 accumulation. SGEMM-cube is not a bit-exact FP32 emulation and does not cover the full FP32 dynamic range. It targets near-FP32 accuracy for inputs representable through FP16 high and residual components.

The analysis supplies explicit constants for this case. When the FP16 high component remains normal and the scaled residual stays within range, the two-word representation retains 22 significand bits, with a representation-error target of $2^{-22}$. The omitted low-low product is bounded entrywise by $u_{16}^2(1+u_{16})^2|A||B|$; its leading term is $2^{-22}|A||B|$, and the exact coefficient is dominated by the classical FP32 accumulation bound for $k\geq5$. Under the stated termwise RN model, Eq.~\ref{eq:main_bound} gives the conservative complete bound, while Eq.~\ref{eq:main_bound_first_order} reduces it to the familiar first-order form $(\gamma_k+14u_{32})|A||B|$. Retaining the higher-order terms places the factor-of-two crossover at $k=15$. This is the two-word FP16/FP32 specialization of the multiword analysis together with the narrow-range scaling principle. The scaling analysis also makes clear that $s_b=12$ is an experimental, range-dependent choice rather than a fixed guarantee over the full FP16 range. Termwise reconstruction is more stable than elementwise accumulation in the evaluated low-exponent regime.

The implementation improves substantially on native FP16 GEMM and is comparable to, and in some tested cases more accurate than, the
OpenBLAS FP32 SGEMM baseline for the evaluated input distributions and exponent range. With L1-aware blocking and double-buffered pipelining adapted to Ascend's software-managed memory hierarchy it reaches 65.3\,TFLOP/s on an Ascend 910A, 77\% of the three-GEMM FP32-equivalent peak, with double buffering alone responsible for a 57\% gain.

Several things remain to be done: a componentwise accuracy study under adversarial inputs, a hierarchical roofline that separates the L1 and L0 bandwidth ceilings, explicit exponent management to widen the supported dynamic range, input-dependent dynamic scaling, and a released implementation. Extending the approach to INT8 and FP8 matrix units is also of interest, though the narrow-range difficulties become considerably more severe there.

\bibliographystyle{elsarticle-num}
\bibliography{SGEMM-cube}

\end{document}